\documentclass{aa}  

\usepackage{graphicx}
\usepackage{txfonts}
\usepackage{booktabs}

\begin{document}

   \title{A modified Milne-Eddington approximation for a qualitative interpretation of chromospheric spectral lines}

\titlerunning{Milne-Eddington approximation for chromospheric spectral lines}
   \author{A.J. Dorantes-Monteagudo
          \inst{1}
          \and
          A.L. Siu-Tapia
          \inst{1}
          \and
          C. Quintero-Noda
          \inst{2,3}
          \and
          D. Orozco Su\'arez
          \inst{1}
          }

   \institute{Instituto de Astrof\'isica de Andaluc\'ia (IAA-CSIC),
              Glorieta de la Astronom\'ia s/n, Apdo.3004, 18008, Granada, Spain
         \and
         Instituto de Astrof\'isica de Canarias (IAC), V\'ia L\'actea s/n, 38205, La Laguna, Tenerife, Spain
         \and 
         Departamento de Astrof\'isica, Univ. de La Laguna (ULL), 38205, La Laguna, Tenerife, Spain
             }

   \date{Received ?; accepted ?}

   \abstract
  {The Milne–Eddington approximation provides an analytic and simple solution to the radiative transfer equation. It can be easily implemented in inversion codes that are used to fit spectro-polarimetric observations to infer average values of the magnetic field vector and the line-of-sight velocity of the solar plasma. However, it is in principle restricted to spectral lines formed under local thermodynamic conditions, that is, to photospheric and optically thin lines.}
  {We show that a simple modification to the Milne-Eddington approximation is sufficient to infer relevant physical parameters from spectral lines that deviate from local thermodynamic equilibrium, as those typically observed in the solar chromosphere.}
  {The Milne-Eddington approximation is modified by including several exponential terms in the source function to reproduce the prototypical shape of chromospheric spectral lines. To check the validity of such approximation, we first study the influence of these new terms on the profile shape by means of the response functions. Then, we test the performance of an inversion code including such modification against the presence of noise. The approximation is also tested with realistic spectral lines generated with the RH numerical radiative transfer code. Finally, we confront the code with synthetic profiles generated from magneto-hydrodynamic simulations carried out with the Bifrost code. For the different tests, we focus on the vector magnetic field and the line-of-sight velocity. The results are compared with the weak field approximation and center of gravity technique, as well.}
  {The response function corresponding to the new terms in the source function have no trade-offs with the response to the different components of the magnetic field vector and line-of-sight velocity. This allows to perform a robust inference of the physical parameters from the interpretation of spectral line shapes. The strategy has been successfully applied to synthetic chromospheric Stokes profiles generated with both standard models and realistic magnetohydrodynamic (MHD) simulations. The magnetic field vector and velocity can be successfully recovered with the modified Milne-Eddington approximation.}  
  {Milne-Eddington model atmospheres including exponential terms are not new to the solar community but have been forgotten for quite some time. We have shown that our modification to the Milne-Eddington approximation succeeds in reproducing the profile shape of two chromospheric spectral lines, namely, the Mg~{\small I}~b2 line and the Ca~{\small II} at 854.2 nm. The results obtained with this approach are in good agreement with the results obtained from the weak field approximation (for magnetic field) and the center of gravity (for velocity). However, the Milne-Eddington approximation possess a great advantage over the classical methods since it is not limited to weak magnetic fields or to a restricted range of velocities.}
   \maketitle
%

\section{Introduction}
 The Milne-Eddington (ME) approximation is a useful method to interpret spectral line radiation under local thermodynamic equilibrium (LTE) and to recover relevant physical information of the solar plasma and magnetic field. It is the solution to the Radiative Transfer Equation (RTE) under the conditions of plane parallel atmosphere, constant magnetic and velocity fields, and a source function with a linear dependency on the opacity. In the Sun, these conditions are typically met in the photosphere and the approximation works properly in this region. Higher up in the chromosphere, the physical conditions start to depart from LTE and therefore they are no longer compatible with the assumptions that lead to the ME approximation.  Departures from LTE conditions lead to non typical spectral line shapes (i.e., non Gaussian-like) and the assumption of a linear source function proposed in the ME approximation is not good enough to reproduce them. 

\cite{1988ApJ...330..493L} presented a modification of the ME approximation proposing a non linear expression for the source function based on the solution of the non LTE (NLTE) transfer problem for a standard two-level atom. The new approximation consists in adding various exponential terms with an explicit dependency on the optical depth to the ME linear source function. These authors showed that the modified ME model with a single exponential term was sufficiently good to reproduce the prototypical Mg~{\small I}~b triplet. They also suggested that the new source function was similar to the original one. 

To test the validity of the new approximation, the authors took a modified semi-empirical umbral model \citep{1986ApJ...306..284M,1987ApJ...318..930L} and considered a constant magnetic field to synthesize the Mg~{\small I}~b triplet. Then, they performed inversions applying different initial model configurations to test the validity of the new approach (i.e., with one and two exponential terms in the  source function) or using either fixed or free thermodynamic parameters. Two remarkable conclusions were drawn from their study: firstly, only one exponential term in the source function was sufficient to reproduce the Mg~{\small I}~b2 profile shape, and secondly, that for successfully retrieving the original model parameters, the inversion should include the four components of the Stokes vector (i.e., the polarization signals) with varying weights among them (i.e, weights of 0.01 for Stokes $I$ and 0.1 for Stokes $Q$, $U$, and $V$). The authors argued that inverting the full Stokes vector with the mentioned weights helped to properly fit the four Stokes profiles simultaneously, because  the trade-offs\footnote{In \cite{2007A&A...462.1137O} the term "trade-offs" was used to refer to correlations between changes in the different physical parameters and their influence on the Stokes parameters.} between the thermodynamic parameters become less relevant. They also showed that there is a correlation between the sensitivity to the magnetic field strength and the line-to-continuum absorption coefficient in the weak field regime. It is important to remark that the authors focused on the line core only and they obtained slightly underestimated field strengths for inclinations around 45$^\circ$.

Based on the idea of \cite{1988ApJ...330..493L}, the main goal of this work is to ascertain whether it is possible to systematically reproduce the shape of chromospheric spectral line profiles using this modified version of the Milne-Eddington approximation (mME, from now on, i.e., ME approximation with a non linear source function). 
In Section 2, we present the mathematical description of the mME approximation, and briefly describe the weak field approximation (WFA) and the center of gravity (CoG) method.
In Section 3, we test the validity of the approximation and its applicability as a fast inversor on synthetic profiles of the Mg I b2 line generated from a FALC model \citep{1993ApJ...406..319F}.
In Section 4, we apply the mME approximation to synthetic profiles generated from a radiative MHD simulation, and compare the retrieved magnetic field and LOS velocity with those obtained from the WFA and CoG, respectively, as well as with the original stratifications in the simulation.  Finally, in Section 5, we discuss the results as well as the capabilities and the shortcomings of the mME approximation.

\section{Methodology} \label{Methodology}
\subsection{Modified Milne-Eddington approximation} \label{mMEsec}
 \begin{figure}[h]
   \centering
   \includegraphics[width=9cm]{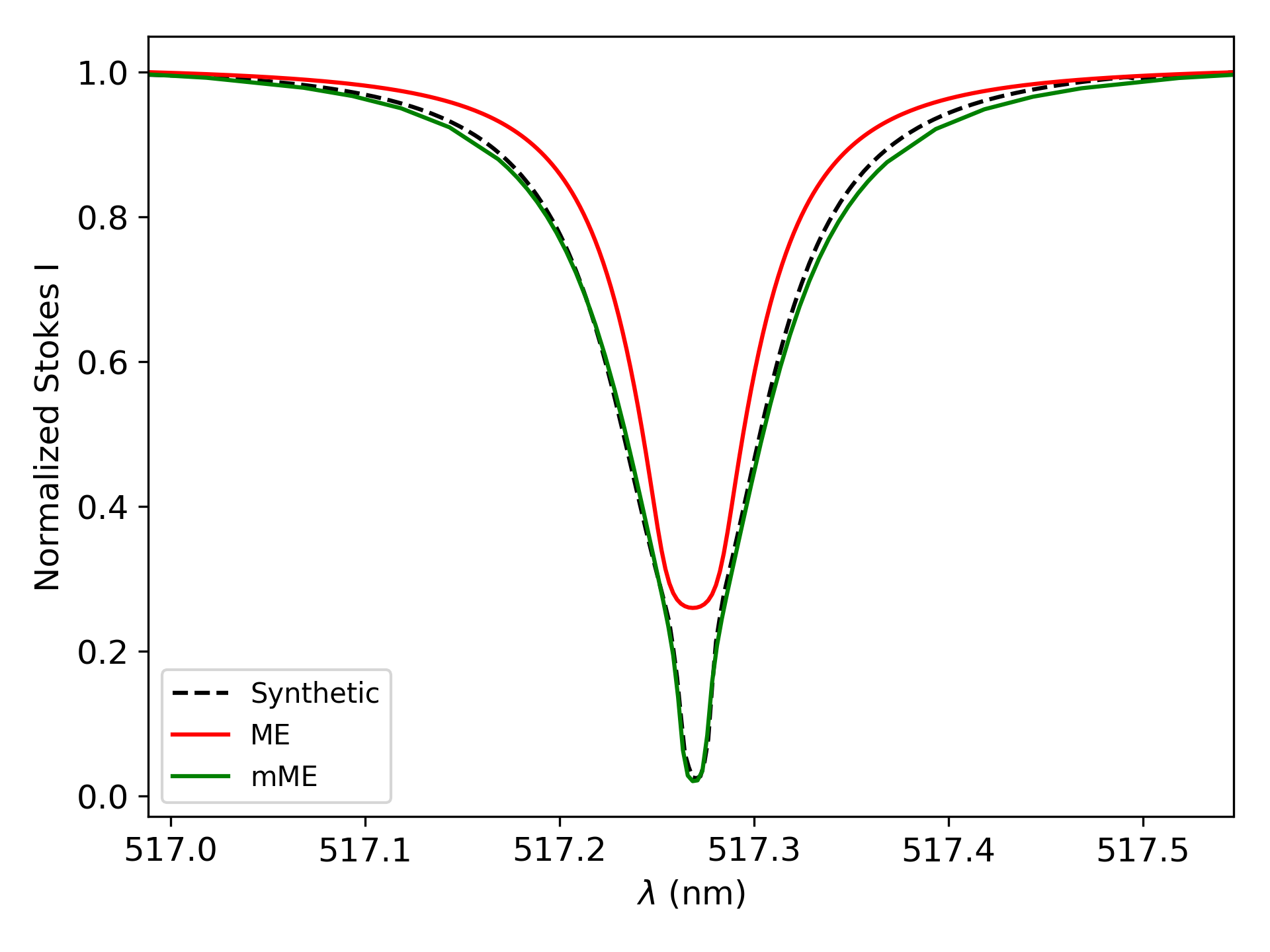}
      \caption{Normalized Stokes $I$ profile corresponding to the Mg~\textsc{I}~b2 spectral line at 517.2~nm. The dashed line corresponds to the synthetic profile generated from MHD simulations (see Sect.~\ref{secsimul}) while the solid red and green lines correspond to the ME and mME fitted profiles, respectively, with model parameters listed in Table~\ref{tab:modelpars}.
              }
         \label{fit}
   \end{figure}
 
The Milne-Eddington approximation to the RTE makes the assumptions that: the atmosphere is plane-parallel and semi-infinite; the physical quantities $B$, $\theta$, $\chi$, and $\Delta\lambda_D$ (magnetic field strength, inclination, azimuth, and Doppler width, respectively) are depth-independent. Consequently, the magnetic field vector, the line-of-sight (LOS) velocity, and the absorption and anomalous dispersion profiles are also independent of the optical depth, $\tau$. Nonetheless, the source function, $S$, depends linearly on $\tau$ as follows:
   \begin{equation}
       S (\tau) = S_0 + S_1 \tau. \label{ME}
   \end{equation}
   Under these assumptions, an analytical solution for the RTE can be found \citep[see, e.g.,][]{1956PASJ....8..108U,1962IzKry..28..259R,1967IzKry..37...56R,1982SoPh...78..355L,2007A&A...462.1137O}. In total, a ME model can be fully described with just nine parameters,  three for the magnetic field vector, one for the LOS velocity, and five for the so-called thermodynamic parameters: $S_0$, $S_1$, the line-to-continuum absorption coefficient ratio, $\eta_0$, the Doppler width, $\Delta\lambda_D$, and the damping parameter, $a$. 
  Therefore, the ME approximation does not provide information about gradients with height of the physical parameters and cannot reproduce asymmetries of the line profiles. Nonetheless, it is a much faster method than typical inversion codes that solve the RTE under LTE assumptions \citep[for a review see][]{2016LRSP...13....4D}, which makes it an ideal method to perform statistical estimates of the magnetic field in the solar photosphere.
However, in order to be applicable in chromospheric spectral lines, the ME approximation requires some modifications.
  
 The chromosphere is the region of the solar atmosphere where the temperature rises, and magnetic pressure starts to dominate over the gas pressure. This leads to a departure from LTE, so to model chromospheric lines it is necessary to solve the RTE in NLTE conditions. 
The ME approximation can be modified in a way so that the source function reproduces the temperature rise in the chromosphere. In practice, this is done by adding depth-dependent exponential terms in the linear source function, as follows:
   \begin{equation}
       S(\tau) = S_0 + S_1 \tau + \sum_{i=1}^{N} A_i e^{-\alpha_i \tau} \simeq S_0 + S_1 \tau + A_1 e^{-\alpha_1 \tau} - A_2 e^{-\alpha_2 \tau}\label{MME}
   \end{equation}
where the new source function has two new parameters, $A$ and $\alpha$, per exponential. To solve the inverse problem and to preserve uniqueness and stability for the solution, one should aim for a low number of exponentials. Fortunately, setting $N=2$ provides two new terms  and four additional free parameters that successfully reproduce the prototypical shape of chromospheric spectral lines. For consistency with \cite{2004ASSL..307.....L}, in Eq.~(\ref{MME}) we have changed the sign in the second exponential. In fact, this new source function yields to new analytical expressions for the Stokes parameters \citep{1988ApJ...330..493L,2004ASSL..307.....L}:
  \begin{equation}
    \begin{split}
        \Vec{I} = \{& S_0\textbf{1}+S_1\textbf{K}^{-1}+A_1[\textbf{1}-\alpha_1(\alpha_1\textbf{1}+\textbf{K})^{-1}]-\\
        &-A_2[\textbf{1}-(1+\alpha_2)(\alpha_2\textbf{1}+\textbf{K})^{-1}]\}\Vec{U} \label{solRTE}
    \end{split}
\end{equation} 
where, $\Vec{I}$ is the Stokes vector whose components are $(I, Q, U, V)^\dagger$; \textbf{K} is the propagation matrix; $\Vec{U}$ is the unit vector $(1,0,0,0)^\dagger$, and $\Vec{1}$ is the identity matrix.

    The role of the two new exponentials is the following: one of them (e.g., $i=1$) is meant to reproduce possible emission features in the line core (through $A_1$) and to make the line wider along the wings ($\alpha_1$). The other one (e.g., $i=2$) narrows the shape of the profiles around the line core ($\alpha_2$) and increases the line-to-continuum contrast ($A_2$). This expression can be reduced to the Unno-Rachkovsky solution when $A_1=A_2= 0$ (i.e., the classical ME approximation for a linear source function). In the case of the mME model  the intensity of the nearby continuum is:
\begin{equation}
    I_c = S_0+S_1 + \frac{A_1}{1+\alpha_1},
\end{equation}
since the line core opacity at the continuum is negligible, $\kappa_L\sim0$, and thus the propagation matrix can be replaced by the identity matrix, $\textbf{K}\sim\textbf{1}$. It is worth to mention that, if $\kappa_L\to\infty$, the intensity of the line core can be written as $I_{core} = S_0+A_1-A_2$. Hence, the mME produce emission in the line core when $A1-A2 < 0$, that is, when $A2 < A1$.

Figure \ref{fit} displays a synthetic Stokes I profile corresponding to the Mg~\textsc{I} b2 line at 517.3~nm, generated from realistic MHD simulations (see Sect.~\ref{secsimul}). The plot compares the fits resulting from both, the ME and the mME approximations. The ME fit resembles a typical Voigt profile and cannot reproduce the synthetic profile. In contrast, the mME approach, with only two exponential terms, is able to reproduce both, the narrow line core and the wide wings. The difference in the quality of the fits between the blue and red wings of the line  illustrates the fact that the synthetic profile is not symmetric while the mME approach provides strictly symmetric Stokes $I$ profiles. Table~\ref{tab:modelpars} summarizes the results obtained for the different free parameters from both inversions. Unlike \cite{1988ApJ...330..493L} who used only the line core, we use a wider spectral window to include the full Mg~\textsc{I} b2 line. The reason is that they used a data set that corresponds to a sunspot, where large magnetic field gradients are expected. In our case, the data set represents a portion of the quiet Sun, where the gradients of the magnetic field with height are milder than in sunspots, and therefore, we can use the full spectral range. This is the reason why they concluded that only one exponential term was enough to reproduce the shape of the line. As we will show later by means of response functions, two exponential terms in the source function are needed to successfully reproduce, not just the core, but the full line profile.

\begin{table}[]
\caption{Model parameters corresponding to ME and mME models.}
\label{tab:modelpars}
\resizebox{0.49\textwidth}{!}{%
\setlength{\tabcolsep}{4pt} 
\begin{tabular}{@{}c|ccccccccc@{}}
    & $S_0$ & $S_1$ & $\eta_0$ & $\Delta\lambda_D$ & $a$ & $A_1$ & $\alpha_1$ & $A_2$ & $\alpha_2$ \\ \midrule
ME  & 0.23 & 0.70 &   12.0       &       0.08            &  0.16   &   -    &   -    &   -   &     -       \\
mME & 0.30 & 0.60 &     90.17     &        0.03           &  0.20   & -0.41 & 0.18  & 0.2  &  98.18         
\end{tabular}%
}
\end{table}

 \subsection{Response Functions in the mME case}
     \begin{figure}[h]
   \centering
   \includegraphics[width=4.45cm]{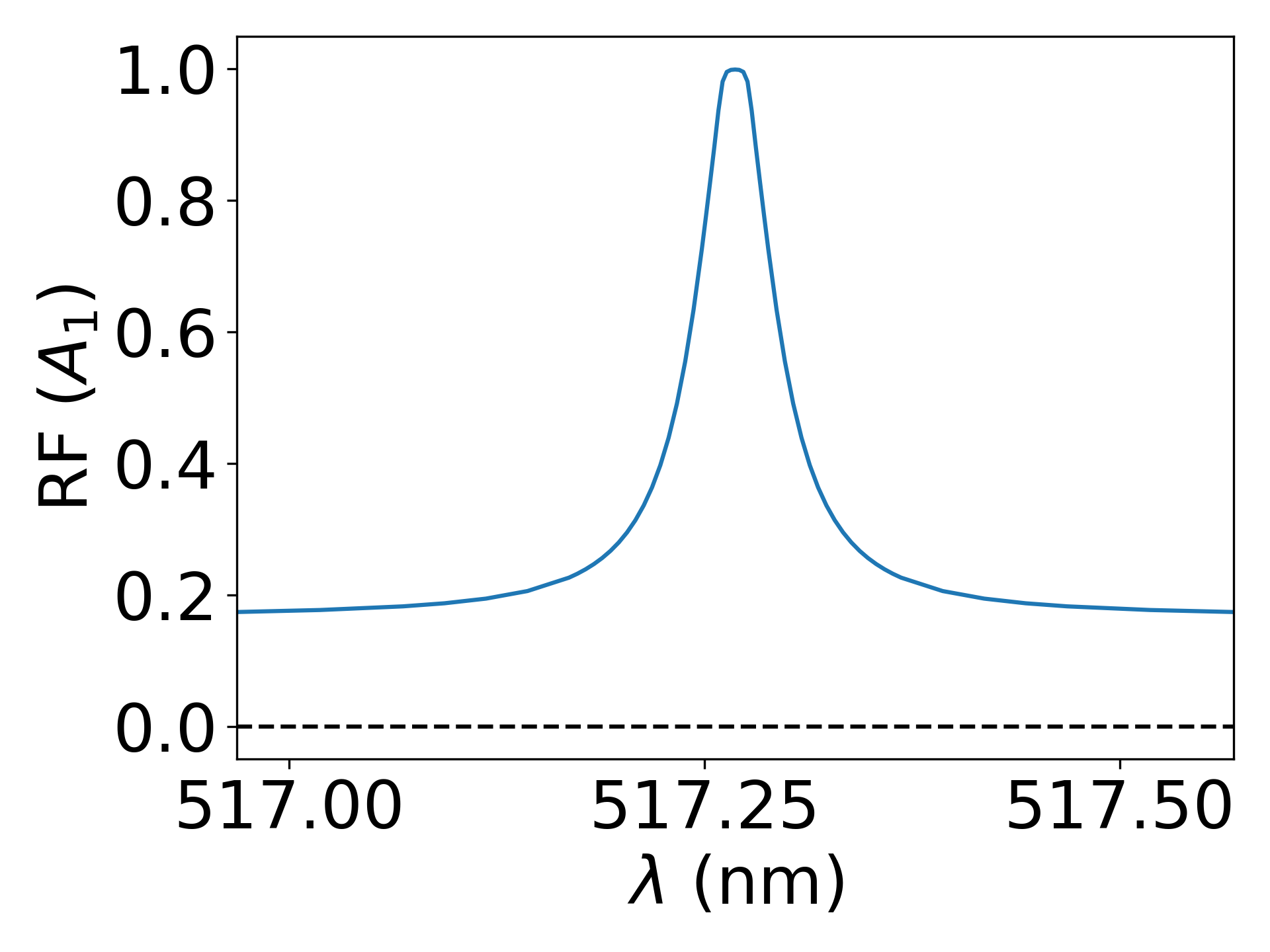}
   \includegraphics[width=4.45cm]{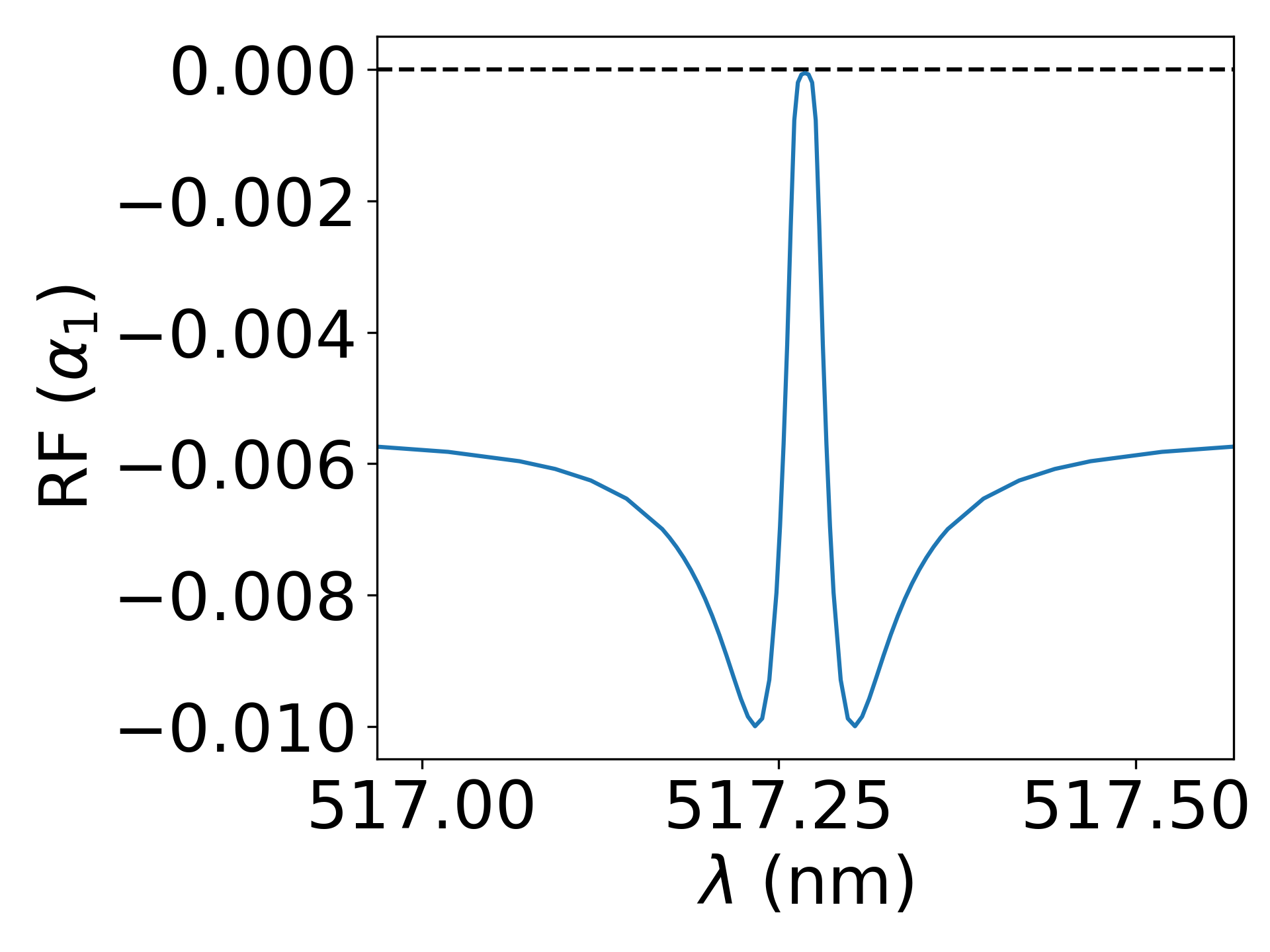}
   \includegraphics[width=4.45cm]{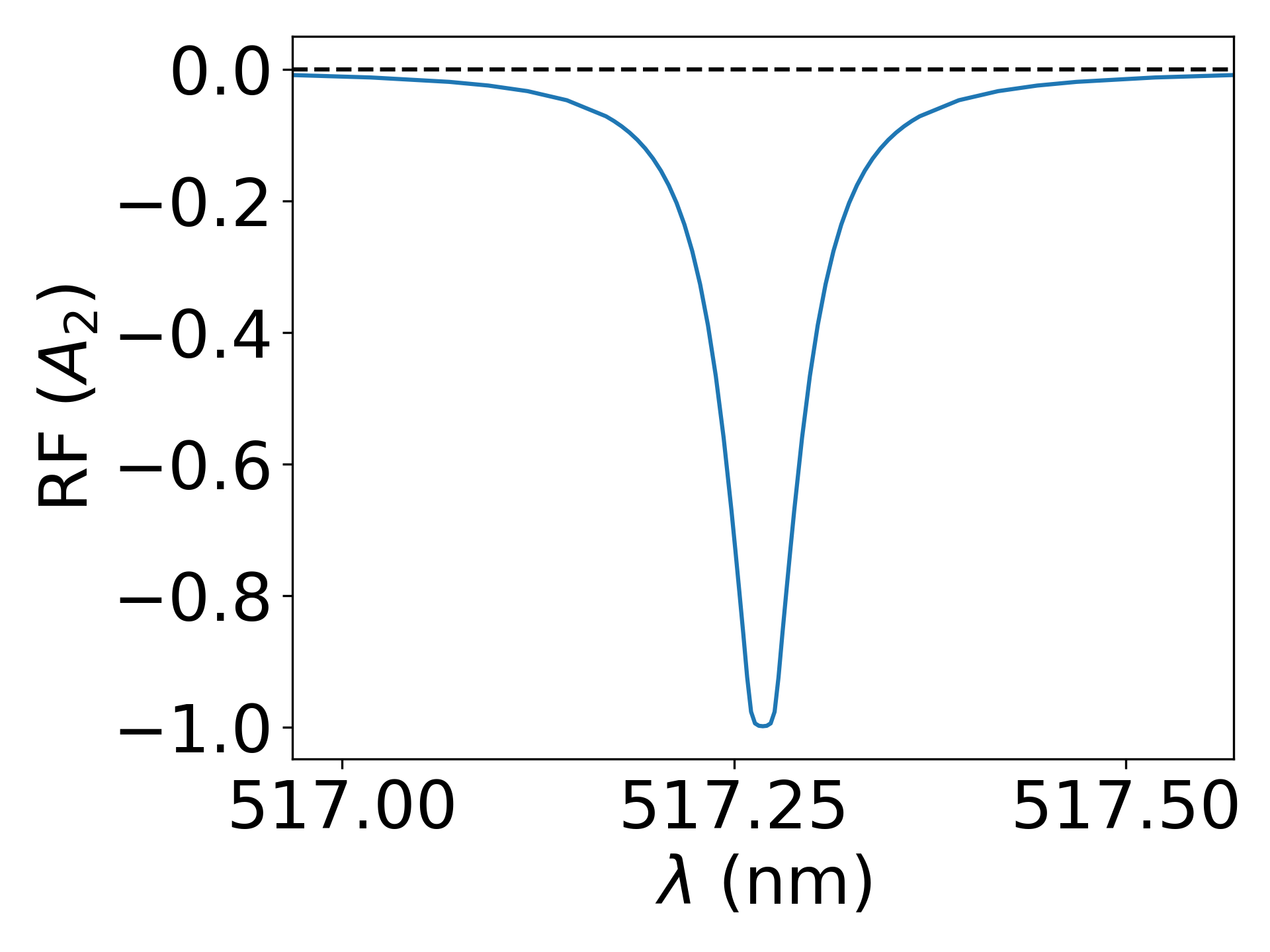}
   \includegraphics[width=4.45cm]{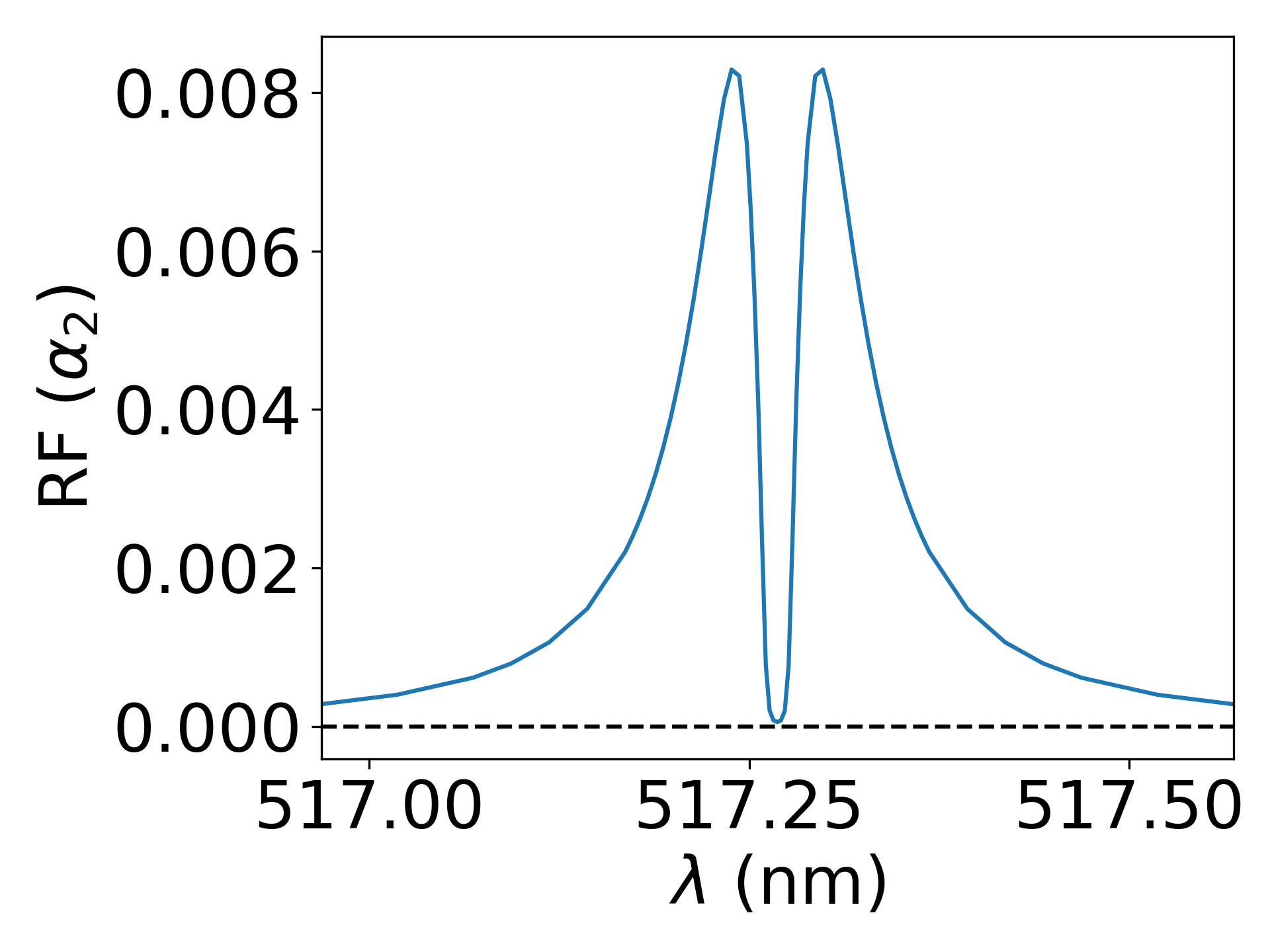}
      \caption{Dimensionless Stokes $I$ Response Functions for the Mg~{\small I}~b2 line adopting the mME fit parameters from Table 1. From left to right and from top to bottom response functions for $A_1$, $\alpha_1$ - Eqs. (\ref{dIA1}), $A_2$ and $\alpha_2$ - Eqs. (\ref{dIA2}), respectively. The dashed line represents zero response.}
         \label{RF}
   \end{figure}
   
 One of the strengths of the ME approximation is that it is possible to derive Response Functions analytically. 
 The response functions give information about how the Stokes profiles change due to the variation of a given physical parameter. They are defined as the partial derivatives of the Stokes profiles with respect to one of the physical parameters in the model \citep{2007A&A...462.1137O}. These authors have shown a very important property of ME model atmospheres through the interpretation of the corresponding RFs: The sensitivity of ME Stokes profiles to perturbations in the magnetic field is uncorrelated with the perturbations in the line-of-sight (LOS) velocity and with the thermodynamic parameters \citep{2010ApJ...711..312D,2016LRSP...13....4D}. This fact makes the ME approach robust when determining those physical parameters. However, there is much more uncertainty in the determination of the thermodynamic parameters since the corresponding response functions show clear trade-offs between them.

 As explained before, the mME approximation with $N=2$ has four new thermodynamic parameters. Hence, describing a mME model requires a total of thirteen free parameters, nine of which are meant just to reproduce the shape of the profiles. The new mME response functions with respect to the new four free parameters can also be calculated analytically. The analytical response functions to the first exponential parameters $A_1$ and $\alpha_1$ are: 
\begin{equation}
     \frac{\partial I}{\partial A_1} = 1-\alpha_1 d_{00}(\alpha_1) \ ~ \  ~ ; \ ~ \  ~ \frac{\partial I}{\partial \alpha_1} = - A_1 d_{00} [1-\alpha_1 d_{00}(\alpha_1)] , \label{dIA1}
 \end{equation}
\begin{equation}
     \frac{\partial Q}{\partial A_1} = -\alpha_1 d_{01}(\alpha_1) \ ~ \  ~ \  ~ \ ~ ; \ ~ \  ~ \frac{\partial Q}{\partial \alpha_1} = A_1 \alpha_1 d_{01}^2(\alpha_1) , \label{dQA1}
 \end{equation}
 \begin{equation}
     \frac{\partial U}{\partial A_1} = -\alpha_1 d_{02}(\alpha_1) \ ~ \  ~ \  ~ \ ~  ; \ ~ \  ~ \frac{\partial U}{\partial \alpha_1} = A_1 \alpha_1 d_{02}^2(\alpha_1) , \label{dUA1}
 \end{equation}
 \begin{equation}
     \frac{\partial V}{\partial A_1} = -\alpha_1 d_{03}(\alpha_1)  \ ~ \  ~ \  ~ \ ~  ; \ ~ \  ~ \frac{\partial V}{\partial \alpha_1} = A_1 \alpha_1 d_{03}^2(\alpha_1) , \label{dVA1}
 \end{equation}
   and to the second exponential parameters $A_2$ and $\alpha_2$:
  \begin{equation*}
     \frac{\partial I}{\partial A_2} = (1+\alpha_2) d_{00}(\alpha_2)-1 ~ \  ;
     \end{equation*}
     \begin{equation}
     \frac{\partial I}{\partial \alpha_2} = A_2 d_{00}(\alpha_2) [1-(1+\alpha_2) d_{00}(\alpha_2)] , \label{dIA2}
 \end{equation}
 \begin{equation}
         \frac{\partial Q}{\partial A_2} = (1+\alpha_2) d_{01}(\alpha_2)  \ ~ \  ~ \  ~ \ ~  ; \ ~ \  ~ \frac{\partial Q}{\partial \alpha_2} = -A_2 (1+\alpha_2) d_{01}^2(\alpha_2) , \label{dQA2}
 \end{equation}
  \begin{equation}
          \frac{\partial U}{\partial A_2} = (1+\alpha_2) d_{02}(\alpha_2) \ ~ \  ~  \  ~ \ ~  ; \ ~ \  ~ \frac{\partial U}{\partial \alpha_2} = -A_2 (1+\alpha_2) d_{02}^2(\alpha_2) , \label{dUA2}
 \end{equation}
  \begin{equation}
         \frac{\partial V}{\partial A_2} = (1+\alpha_2) d_{03}(\alpha_2) \ ~ \  ~ \  ~ \ ~  ; \ ~ \  ~ \frac{\partial V}{\partial \alpha_2} = -A_2 (1+\alpha_2) d_{03}^2(\alpha_2) , \label{dVA2}
 \end{equation}
 where $d_{0i}$ are the elements of the inverse matrix defined by $\textbf{D}(\alpha) = (\alpha \textbf{1} + \textbf{K})^{-1}$ which are given by
   \begin{equation*}
   \begin{split}
      d_{00}(\alpha) = \Delta^{-1}(\alpha)(1+\alpha+k_I)[(1+\alpha+k_I)^2+f_Q^2+f_U^2+f_V^2],
   \end{split}
         \end{equation*}
      \begin{equation*}
      \begin{split}
      d_{01} (\alpha) =  -\Delta^{-1}(\alpha)[&(1+\alpha+k_I)^2k_Q+(1+\alpha+k_I)(k_Uf_V-k_Vf_U)  \\ &+f_Q(k_Qf_Q+k_Uf_U+k_Vf_V)],
   \end{split}
      \end{equation*}
      \begin{equation*}
      \begin{split}
      d_{02}(\alpha) = -\Delta^{-1}(\alpha)[&(1+\alpha+k_I)^2k_U+(1+\alpha+k_I)(k_Vf_Q-k_Qf_V)\\ &+f_U(k_Qf_Q+k_Uf_U+k_Vf_V)],
      \end{split}
      \end{equation*}
      \begin{equation}
      \begin{split}
      d_{03}(\alpha) = -\Delta^{-1}(\alpha)[&(1+\alpha+k_I)^2k_V+(1+\alpha+k_I)(k_Qf_U-k_Uf_Q)\\ &+f_V(k_Qf_Q+k_Uf_U+k_Vf_V)],
      \end{split} \label{inverse}
   \end{equation}
   according to \cite{2004ASSL..307.....L}. The term $\Delta(\alpha)$ denotes the determinant of the matrix  \textbf{D}.
   \begin{equation}
       \begin{split}
           \Delta(\alpha) =  (1+\alpha & +k_I)^4 + (1+\alpha+k_I)^2(f_Q^2 +f_U^2 + f_V^2 \\ & -k_Q^2-k_U^2-k_V^2) -(k_Qf_Q+k_Uf_U+k_Vf_V). 
       \end{split} \label{determ}
   \end{equation}
   In the mME model atmosphere there are more chances for trade-offs between the thermodynamic parameters and the vector magnetic field and LOS velocity. Fortunately, the sensitivity of the Stokes profiles to these new parameters show no trade-offs with  the magnetic field vector or the LOS velocity. Figure~\ref{RF} shows the analytical Stokes $I$ response functions to the four parameters describing the exponential terms. The shape of the response functions shows how these parameters modify the profile shape to accommodate the prototypical chromospheric line profile shapes, that is, the possible emission features in the line core and the wider wings and narrower line core. By comparing these response functions with the ones presented in \cite{2007A&A...462.1137O} one can easily check that there are no trade-offs between the thermodynamic parameters and the vector magnetic field and the velocity. Although there is a strong trade-off between the different thermodynamic parameters as in the ME case. 
 
 For the tests described in this paper, we have numerically implemented the mME model in the LMpyMilne (LMfit Inversion in a Milne-Eddignton atmosphere) code\footnote{The code has been developed by C. Diaz Baso and can be found in https://github.com/cdiazbas/LMpyMilne and written in Python language} and in the MILOS\footnote{Available in IDL (Interactive Data Language) at https://github.com/vivivum/MilosIDL} code \citep{2007A&A...462.1137O}. In particular, we added two exponential terms to the source function and used the new analytical solution to the RTE - Eq. (\ref{solRTE}). Both codes are based on the Levenberg-Marquardt algorithm for minimizing the residuals of the Stokes profiles in order to fit them. During the inversion process, all the parameters are defined to be always positive except for $A_1,A_2$, and $v_{los}$.

\subsection{The weak field approximation method}\label{WFAsec}
We  apply the weak field approximation (WFA) to our data and compare the results with our method. When the Zeeman splitting in the spectral line is much smaller than the Doppler width of the line ($\Delta\lambda_B<<\Delta\lambda_D$) one can perform a perturbative analysis of the RTE \citep{1973SoPh...31..299L,1989ApJ...343..920J, 2004ASSL..307.....L, 2018ApJ...866...89C} and deduce some properties of the solutions to the RTE without solving it formally.  The first order approximation leads to an expression that relates the circular polarization (Stokes $V$) with the derivative of the intensity (Stokes $I$):
\begin{equation}
    V(\lambda) = -\Delta\lambda_B \bar{g} \cos\theta \frac{\partial I(\lambda)}{\partial \lambda} \label{WFAcircP}
\end{equation}
where $\bar{g}$ the effective Land\'e factor of the transition. The contribution from the magnetic field to this expression comes from the LOS component by means of $\Delta\lambda_B = k \lambda_0^2 B$ (where $k=4.6686\times10^{-13} \text{G}^{-1}$ \AA$^{-1}$) and $\cos\theta$, with $\theta$ the inclination of the magnetic field and $\lambda_0$ the rest wavelength of the spectral line of interest.

 The linear polarization profiles, $Q$ and $U$, are given by the second order approximation by considering the azimuth as constant and equal to zero, which leads to:
\begin{equation}
    Q(\lambda) = -\frac{1}{4}\Delta\lambda^2_B\bar{G} \sin^2\theta \frac{\partial^2 I(\lambda)}{\partial^2 \lambda},  \label{WFAlinearP}
\end{equation}
where $\bar{G}$ is the transversal Landé factor, defined as:
\begin{equation}
       \overline{G} = \bar{g}^2-\delta \label{GG}
   \end{equation}
   with
   \begin{eqnarray}
       \delta = \frac{1}{80}(g_u-g_l)^2\{16 [J_u(J_u+1)+J_l(J_l+1)]\nonumber\\
       -7[J_u(J_u+1)-J_l(J_l+1)]^2-4\}, \label{delt}
   \end{eqnarray}
where $J_u$ and $J_l$ being the total angular momentum of the upper and lower level, respectively, and $g_u$ and $g_l$ are Land\'e factors of these levels. Strictly speaking, the WFA approximation shown in Eq.~(\ref{WFAlinearP}) is only valid at the line center \citep{2004ASSL..307.....L}. However, since our goal is to make a one-to-one comparison with the mME model we have decided to apply the WFA as such to the whole line profile.

From the boundary conditions it can be deduced that $U(\lambda)/Q(\lambda)=\tan2\chi$ for $Q(\lambda)\neq0$, so this expression is valid in the reference frame where  $U(\lambda)=0$. Using these expressions one can infer the longitudinal and transverse magnetic field components by performing a minimization of the residuals between the observed circular and linear polarization signals and the derivative of the intensity. The minimization can be obtained analytically and leads to the following expression for the longitudinal component of the magnetic field with respect to the Line of Sight (LOS, from now on):
\begin{equation}
    B_\parallel = -\frac{\sum_i V_i \left(\frac{\partial I}{\partial \lambda}\right)_i}{C_1 \sum_i\left(\frac{\partial I}{\partial \lambda}\right)^2_i} \label{parB}
\end{equation}
where $C_1=k\bar{g}\lambda_0^2$. 
Likewise, using Eq. (\ref{WFAlinearP}) and following the same procedure than before, but now relating linear polarization signals (i.e.,~ Stokes $Q$ and $U$) with the second derivative of Stokes $I$, one obtains an expression for the so-called transverse magnetic field component with respect to the LOS:
\begin{equation}
       B_\perp = \sqrt{\frac{\sum_i L_i \left|\frac{\partial^2I}{\partial \lambda^2}\right|_i}{C_2 \sum_i \left|\frac{\partial^2I}{\partial\lambda^2}\right|^2_i}} \label{horB},
   \end{equation}
   where $C_2$ is a constant defined as $C_2 = k^2\lambda_0^4\overline{G}/4$ and $L_i=\sqrt{Q_i^2+U_i^2}$ is the total linear polarization, which in our reference frame is given only by Stokes $Q$ since $U=0$.

\subsection{The center of gravity method}\label{CoGsec}

Following \cite{2003ApJ...592.1225U} and references therein \citep[e.g., ][]{1967AnAp...30..513S}, the LOS velocity can be determined as 
\begin{equation}
    v_{CoG} = \frac{c (\lambda_0-\lambda_{CoG})}{\lambda_0} \label{vCoG},
\end{equation}
with $\lambda_{CoG}$ defined as
\begin{equation}
    \lambda_{CoG} = \frac{\int \lambda (I_{c}-I) d\lambda}{\int (I_{c}-I) d\lambda} , \label{CoG}
\end{equation}
 where $I_{c}$ is the continuum intensity; $I$ the intensity at a given wavelength, $\lambda$; $\lambda_{CoG}$ the center of gravity wavelength; and $c$ the speed of light.
   
   \section{Validity of the mME model} \label{Validity}
To test the validity of the mME approximation in chromospheric spectral lines, we consider different scenarios. We first check the convergence and uniqueness of the solution using mME profiles (Sect.~\ref{cuniq}). Later, we generate realistic Stokes parameters with the RH NLTE synthesis code \citep{2001ApJ...557..389U, 2003ApJ...592.1225U}. We performed the synthesis in a wavelength range that goes from -400 to +400 m\AA{} around the line core, with a sampling of 10 m\AA{}. We did not take into account PRD effects because they are negligible for Mg~{\small I}~b2, so it is performed under CRD conditions \citep{2018MNRAS.481.5675Q} (Sect.~\ref{cmodels}).

\subsection{Convergence and uniqueness of the solution against noise}
\label{cuniq}
\begin{figure}[!http]
   \centering
   \includegraphics[width=9cm]{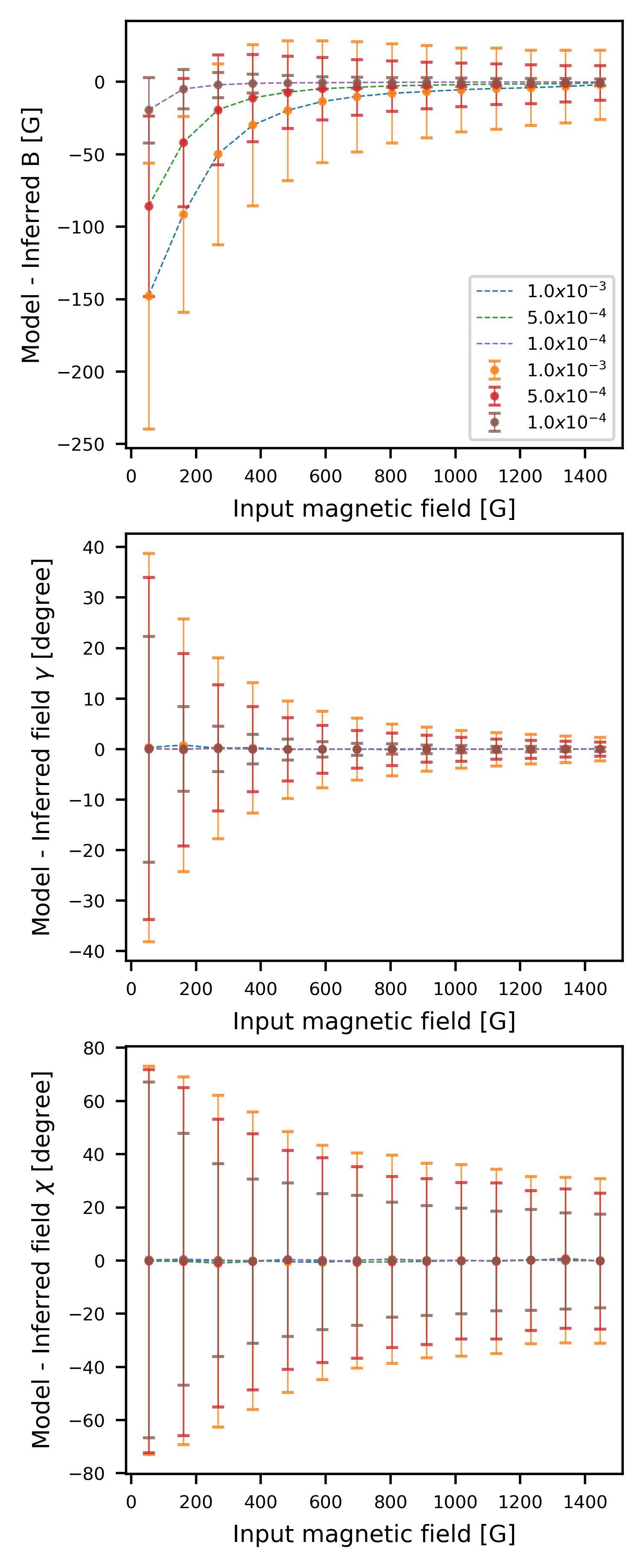}
      \caption{Errors for the magnetic field strength, field inclination, and azimuth resulting from the inversion of $100\,000$ Stokes profiles with different noise levels. The dashed lines show the mean value of the departure from the input, and the error bars indicate the standard deviations.}
         \label{robustness}
   \end{figure}
Although the mME approximation can be written down analytically, the extraction of the physical quantities from the observed Stokes profiles requires the use of inversion techniques, that is, an automatic procedure that minimizes the squared differences between a synthetic Stokes profile (associated to a given model atmosphere) and the observed one. The absence of trade-offs between the different model parameters provides confidence for the determination of the different physical quantities. However, the inversion algorithm can also introduce uncertainties into the process. For instance, the algorithm might not be robust enough due to the large number of parameters to be fit, or due to the noise in the observations. Hence, to check the robustness of the inversion code, a set of synthesized Stokes profiles emerging from 100 000 mME model atmospheres with added random noise at the level of $10^{-3} I_c$, $0.5\times 10^{-3} I_c$, and $10^{-4} I_c$ were generated assuming a random model atmosphere. The model was built with an uniform random distribution of vector magnetic fields (with strength from 0 to 1500~G and inclination and azimuth from 0$^{\circ}$ to 180$^{\circ}$) and LOS velocities between $-2$~km s$^{-1}$ to $+2$~km s$^{-1}$. Both the magnetic field vector and the LOS velocity are constant with height. The thermodynamic parameters are allowed to vary randomly around 20\% of the initial value. The initial model parameters were: $S_0 = 0.06$, $S_1=0.86$, $\eta_0 = 900$, $\Delta\lambda_D = 0.56$, $a = 0.03$, $A_1 = 0.74$, $\alpha_1 = 11.42$, $A_2 = 0.76$, and $\alpha_2 = 25.58$. All the tests were done with the Mg~{\small I}~b2 line at 517.3~nm so the thermodynamic parameters correspond to those that best reproduce this spectral line. The number of spectral samples across the line was $100$ with a sampling of 0.3~pm. In the inversion we allowed a maximum of 100 iterations and, contrary to \cite{1988ApJ...330..493L}, we did not need to apply different weights. We think variable weights, as they established, are really necessary as soon as the Stokes profiles show asymmetries. The averaged time of the inversion of those random pixels is $\sim 0.6$ s per pixel.

The inversion results can be seen in Fig.~\ref{robustness}. The root mean squared (rms) errors depend on the amount of noise applied to the simulated profiles. In detail, for magnetic fields stronger than 200~Gauss the rms errors of the magnetic field strength are $\approx 37$, $19$, and  $4$~G for $10^{-3} I_c$, $0.5\times 10^{-3} I_c$, and $10^{-4} I_c$ noise levels, respectively. For weaker fields, the rms values are $\approx 80$, $53$, and $18$~G, although the most important contribution to the error comes from a clear deviation of the mean value for the inferred field strengths. In particular, the field strength is always overestimated by more than $100$~G when the field is smaller than $200$~G for a noise level of $10^{-3} I_c$. The deviation is almost negligible as soon as the Stokes profiles are affected by an rms noise of $10^{-4} I_c$. For the field inclination, the noise increases monotonically as the field strength decreases, being on average $6.5^{\circ}$, $4.2^{\circ}$, and $1.4^{\circ}$ for $10^{-3} I_c$, $0.5\times 10^{-3} I_c$, and $10^{-4} I_c$ noise levels, respectively, and for fields stronger than $200$~G. The azimuth is the most affected parameter by the noise, with rms values always above $20$~degrees.

\subsection{Comparison with standard chromospheric models} \label{cmodels}

     \begin{figure}[!http]
   \centering
   \includegraphics[width=9cm]{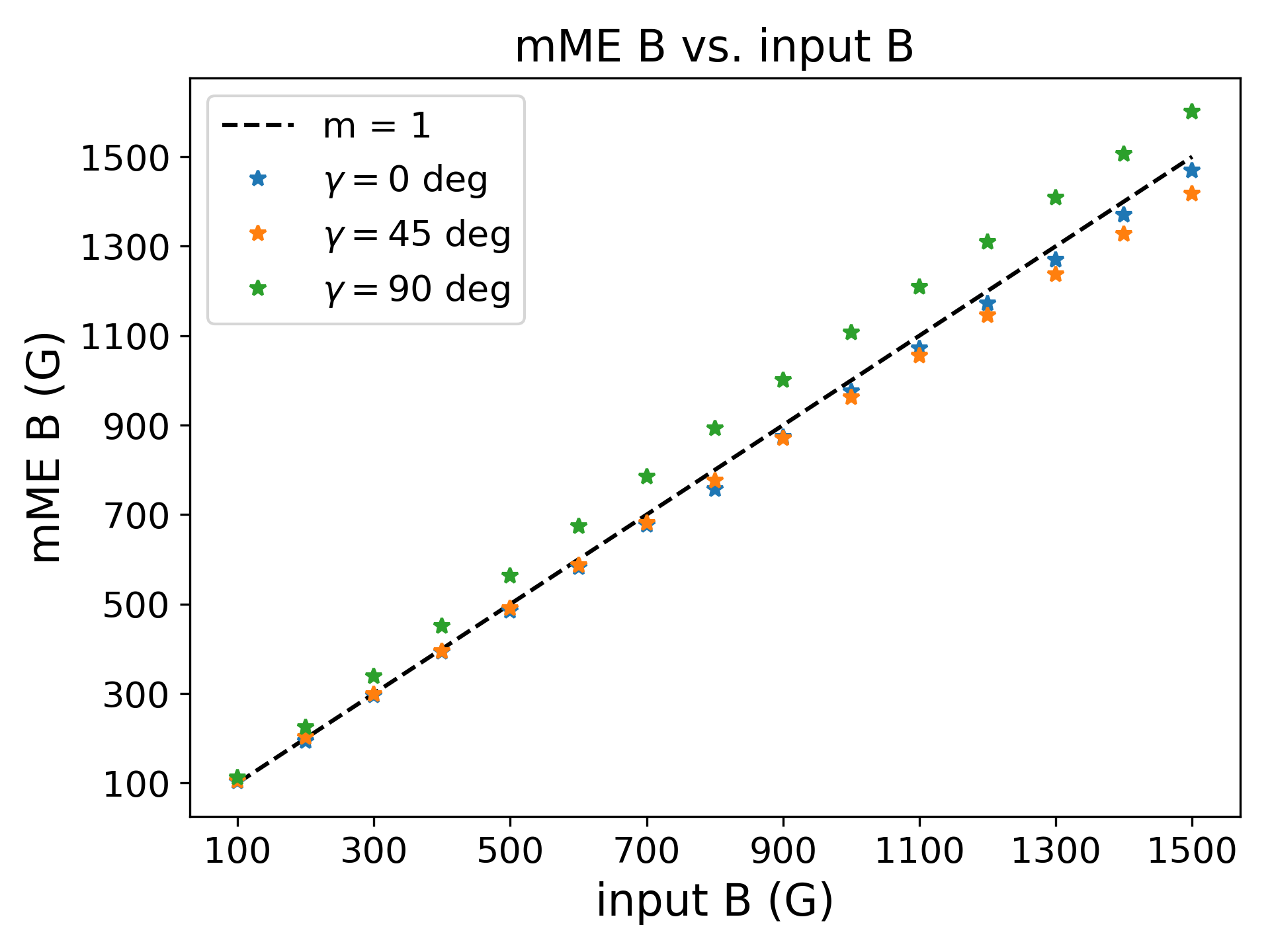}
   \includegraphics[width=9cm]{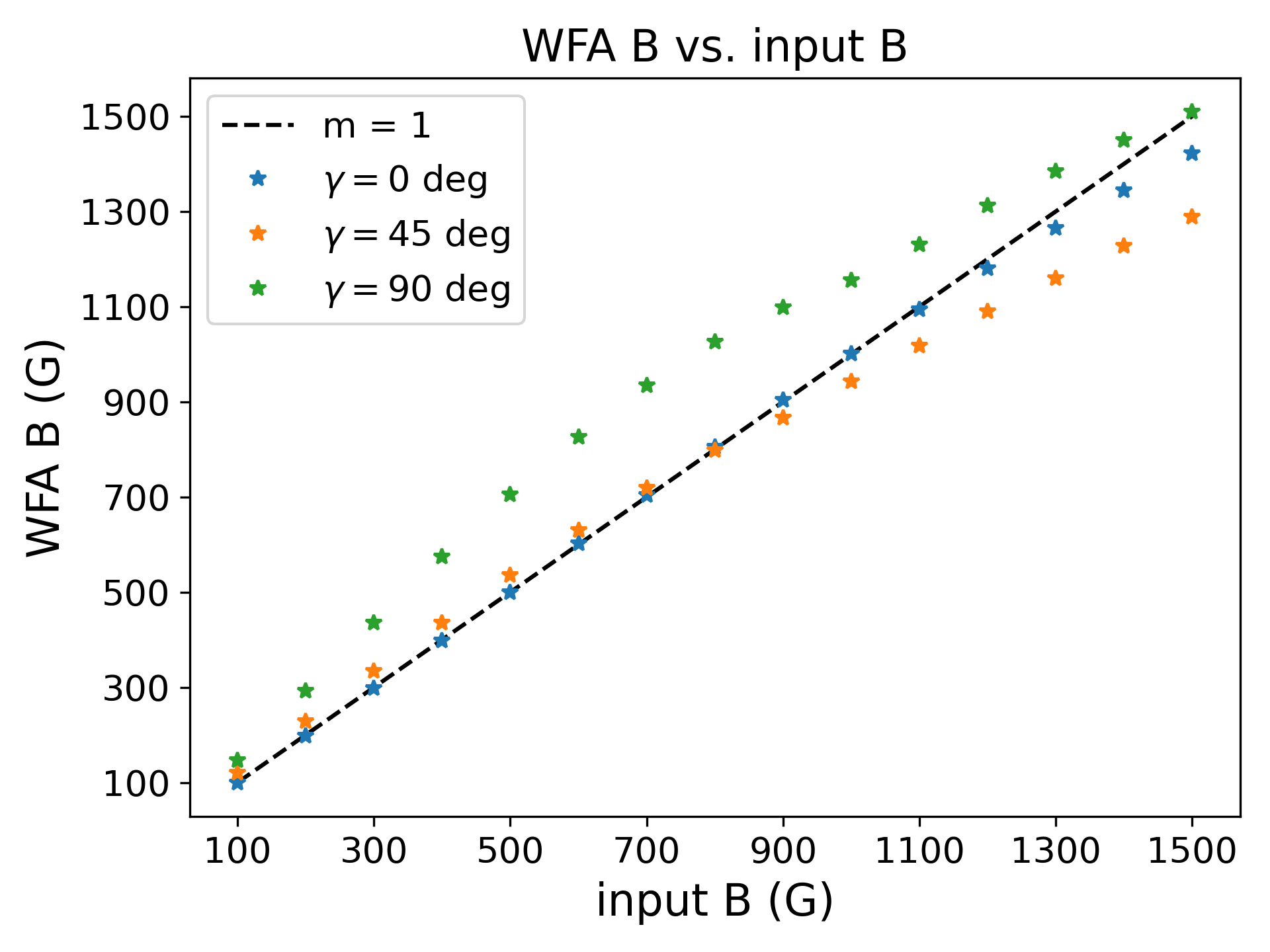}
      \caption{Inferred vs.\ input magnetic field strengths corresponding to the mME approximation (top) and to the WFA (bottom), for different field inclinations, $\gamma$. The black dashed line is the identity line.}
         \label{mMEvsreal2}
   \end{figure}

In this subsection we test the robustness of the mME approximation against realistic Stokes profiles. In particular, we have combined different magnetic field configurations with a FALC model atmosphere \citep{1993ApJ...406..319F} and generated a set of Stokes profiles with the RH code. In the models, the magnetic field strength spans from 100 G to 1500 G in steps of 100 G while the field inclination is set to 0$^\mathrm{o}$, 45$^\mathrm{o}$, and  90$^\mathrm{o}$. In all models, the magnetic field azimuth is constant and set to zero degrees. For the tests, we have concentrated on the Mg~{\small I}~b2 line at 517.2~nm. We have performed 100 mME inversions for each of the magnetic field configurations mentioned above. We use the same initialization each time but adding a varying random noise signal of $0.5\times 10^{-3} I_c$ to the synthetic Stokes profiles. We have also applied the WFA in different field regimes: weak, intermediate, and strong fields.

Figure \ref{mMEvsreal2} shows the mean magnetic field strength that results from the 100 fits for each model using the mME approximation (top panel) and the WFA (bottom panel), both versus the input magnetic field strength values.
In the mME case, the inferred field strengths are slightly overestimated for horizontal magnetic fields (90$^\mathrm{o}$) and the deviations to the input values, in terms of relative errors, reach up to 13\% for the weakest fields and about 6.7\% for the strongest ones; while for intermediate inclinations (45$^\circ$) and longitudinal fields (0$^\mathrm{o}$) the resultant magnetic fields are slightly underestimated. This behavior was already pointed out by  \cite{1988ApJ...330..493L}. However, for 45$^\circ$ inclinations they obtained errors of about 9\% after including the full Stokes vector in the inversion, while in our case, the inferred values are underestimated up to 3\% but such relative errors do not show a clear dependency on the magnetic field strength. The smaller error values of the retrieved magnetic field strengths may be simply because we consider a larger spectral window than \cite{1988ApJ...330..493L}. It is important to remark that the standard deviations calculated from the 100 inversions are very small (of the order of 1 G), which means that the deviations from the identity line are not due to poor fits. 
 
The bottom panel in Fig. \ref{mMEvsreal2} shows the limitations of the WFA when it is applied to realistic Stokes profiles. It can be seen how the  inferred fields significantly deviate from the input fields at inclinations of 0$^\circ$ and 45$^\circ$  for strengths above 1000~G. This is because the WFA is valid  for magnetic fields where the Zeeman splitting is smaller than the Doppler width of the line. Therefore, the stronger the fields, the larger the errors are. Anyhow, for vertical fields, the deviation of the WFA is slightly smaller than in the mME approximation. However, it can also be noticed that errors get significantly larger as the magnetic field inclination increases, in part because the transverse component of the field is based on a second order approximation, unlike the longitudinal component of the magnetic field. 

It can be argued that the mME approximation in general provides better results than the WFA for different magnetic field strengths and inclinations. It has limitations though, as the deviations may be as large as 200~G for horizontal fields. 

 \subsection{Source function determination}

 \begin{figure}[h]
   \centering
      \includegraphics[width=9cm]{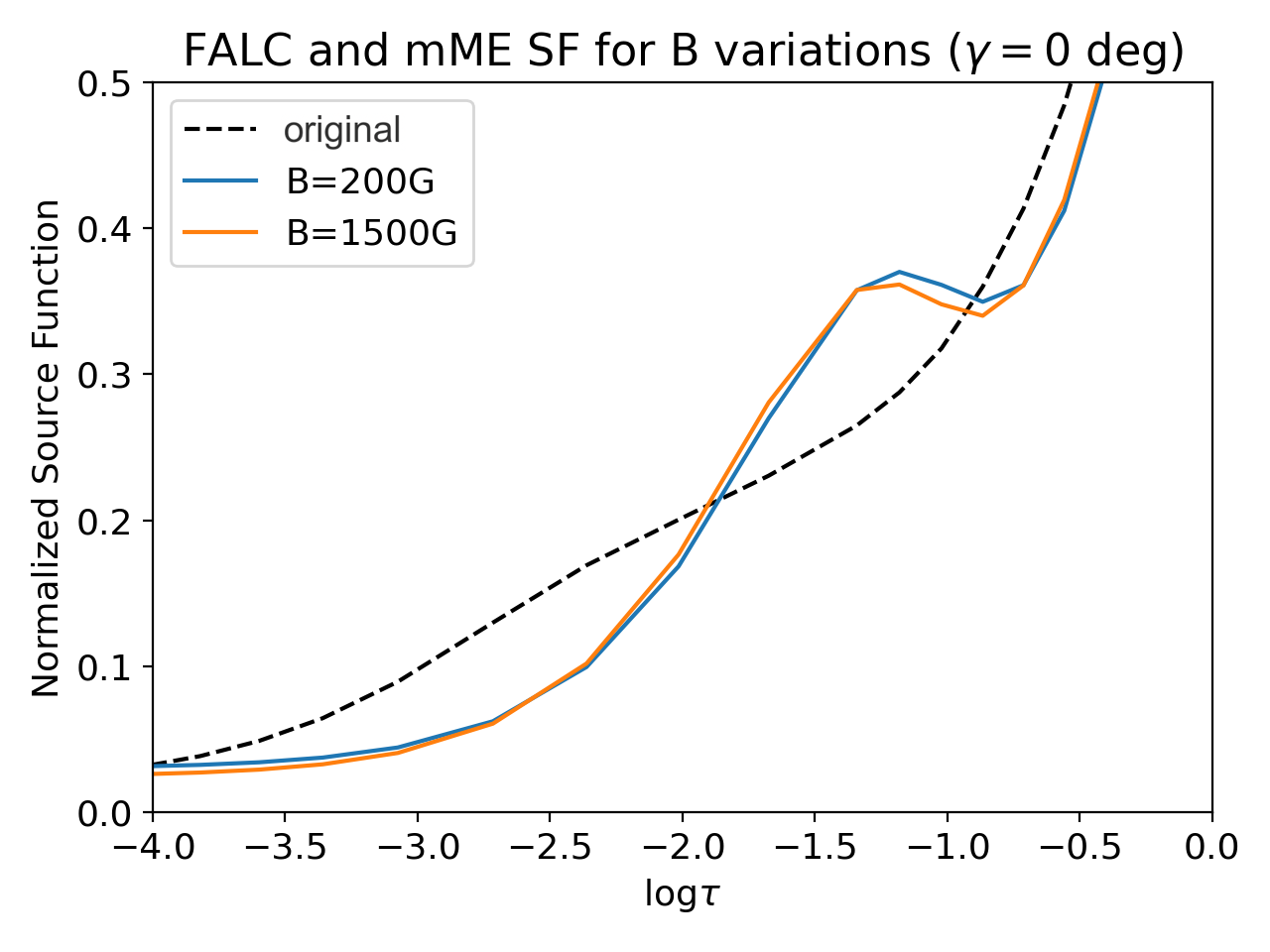}
      \includegraphics[width=9cm]{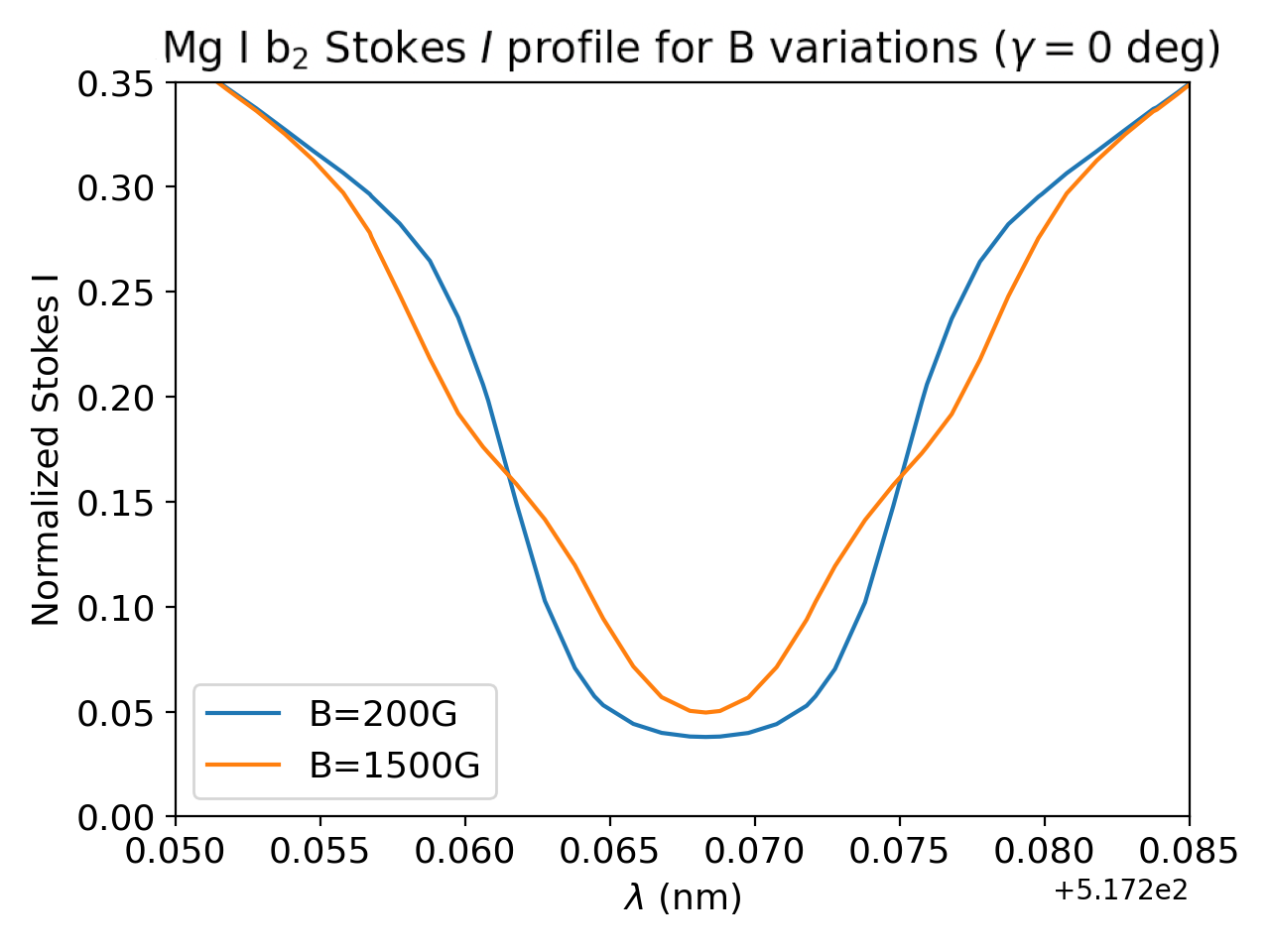}
      \caption{mME inferred source functions (top) and associated intensity profiles (bottom) corresponding to the lowest (200 G; blue line) and the highest (1500 G; orange line) values of the magnetic field strengths added to the FALC model atmosphere. Dashed line in the top panel corresponds to the original source function calculated with the RH code. The bottom panel only shows the line core. 
      }
         \label{SF4B}
   \end{figure}
   
      \begin{figure*}
   \centering
   \includegraphics[width=19cm]{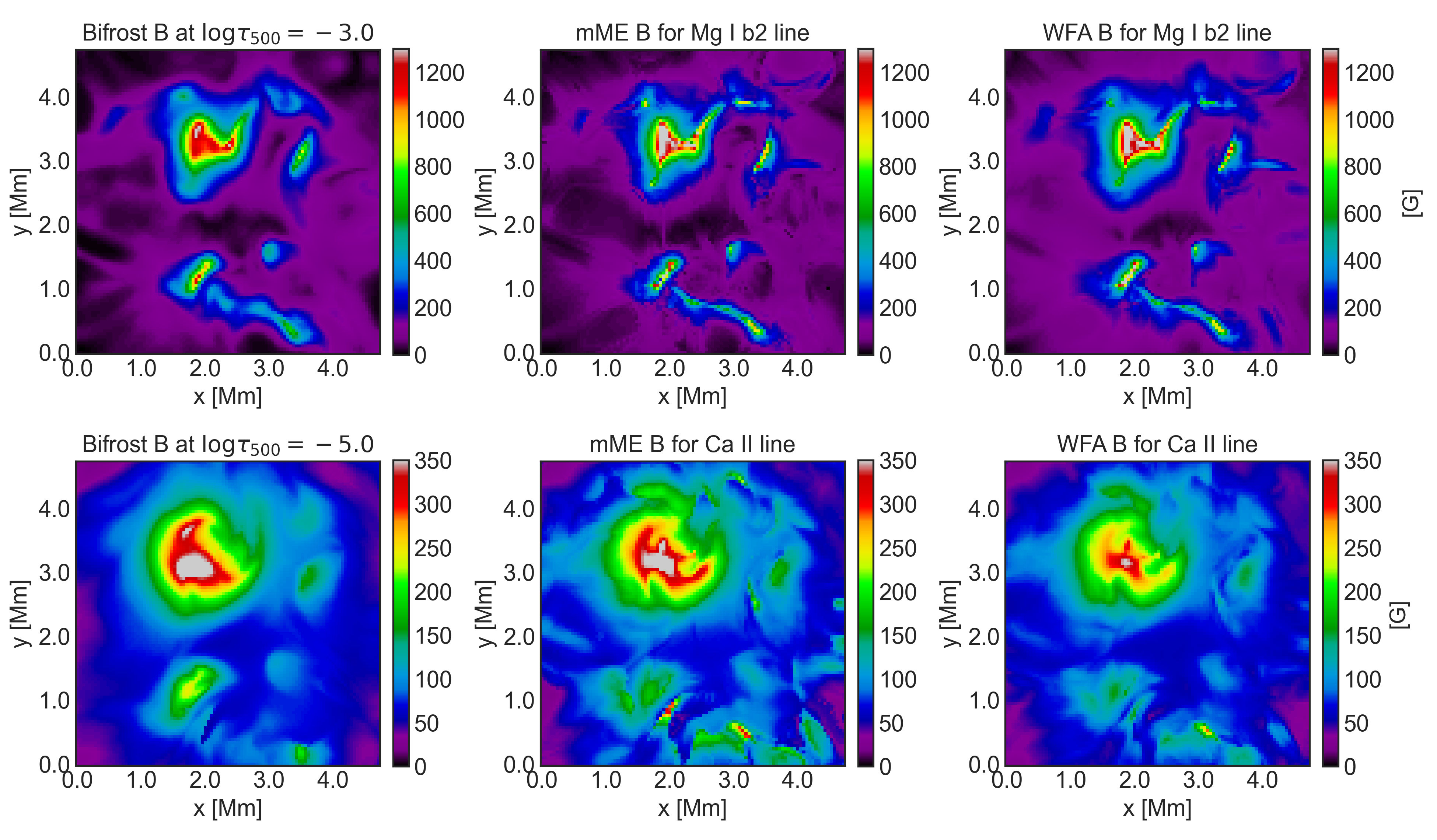}
   \caption{Maps of the magnetic field strength in the region of interest from the Bifrost simulation (left panels), from the mME approximation (central panels), and from the WFA (right panels); and for the Mg~{\small I}~b2 line (top) and for the Ca~{\small II}~854.2~nm (bottom). The maps in the left panels are taken at the estimated optical depths according to the mME inversion results ($\log\tau_{500}=-3.0$ for Mg~{\small I}~b2 and $\log\tau_{500}=-5.0$ for Ca~{\small II}).}
              \label{BB}%
    \end{figure*}

\cite{1988ApJ...330..493L} pointed out that the mME source function they obtained was quite similar to the original one, which might open the possibility to infer temperatures from the mME source functions. However, we have found that there are clear discrepancies between both source functions (see Fig.~\ref{SF4B}). The reason is that the mME approximation considers a constant propagation matrix (\textbf{K}) with optical depth, which would certainly not be enough for reproducing chromospheric profile shapes. The mME uses the exponential terms of the source function to reproduce the line shapes discarding the physics of the line formation. This is the reason why the mME source function deviates from the real one. Nonetheless, the approximation can successfully reproduce the line profile shapes, provided they are symmetric. 
 
Fig.~\ref{SF4B} shows a comparison between the source function inferred with the mME model for two different magnetic field strengths and the original source function computed by the RH synthesis code. The mME source function is understood as a \emph{thermodynamic parameter} and, as discussed in the previous section, it does not show trade-offs with the parameters describing the vector magnetic field. Hence, the two results are quite similar except for a small deviation at around $\log\tau = -1.0$. However, the source function is quite different except from the fact that all of them tend to decrease towards higher atmospheric layers, for example, there are clear humps around $\log\tau = -1.2$ in the mME source functions, while the source function shows the hump at around $\log\tau = -2.5$ with a much smaller amplitude.

\section{Comparison against realistic MHD models} 
\label{secsimul}
We use the 3D radiative magneto-hydrodynamic simulation of a portion of the enhanced network run performed with the \emph{Bifrost} code \citep{2011A&A...531A.154G} by \cite{2016A&A...585A...4C}. The physical size of the original box is 24 Mm$\times$24 Mm$\times$16.8 Mm with a 48 km sampling in the horizontal domain and a 19 km sampling in the vertical one. The model goes from 2.4 Mm below the photosphere (upper convection zone) to about 14.4 Mm above it, including the chromosphere and part of the corona.

The magnetic field in the simulation is a bipolar structure seen at photospheric heights as two clusters of magnetic field concentrations that have similar strength but display opposite polarities. Here, we focus on a 100$\times$100 pixels region of positive polarity. The area contains pixels with both strong and weak magnetic field concentrations, so that the mME approximation can be tested for a wide range of magnetic field strengths. From these simulations, \cite{2018MNRAS.481.5675Q} synthesized the Mg~{\small I}~b triplet (b1, b2 and b4) and the Ca~{\small II}~854.2~nm line with a wavelength sampling of 10 m\AA\/ using the RH code. Here we use those synthetic profiles in the cropped region.

We have inferred the line-of-sight velocity and the magnetic field vector from the synthetic profiles using the mME approximation. The mME inversion has been performed in a pixel by pixel basis, hence obtaining for each of the pixels the magnetic field strength ($B$), inclination ($\gamma$), azimuth ($\phi$), the LoS velocity ($v_{LOS}$), the line-to-continuum absorption coefficient ratio ($\eta_0$), the damping coefficient ($a$), the Doppler width ($\Delta\lambda_D$), and the six parameters describing the mME source function (Sect.~ \ref{mMEsec}). The inversion was repeated three times: first, the initial model was initialized randomly; after the first inversion, we calculated the mean value of each of the inferred model parameters and set them as the initial model for a second inversion; finally, the process is repeated using the results of the second inversion. The first tests showed that if the  thermodynamic parameters in the input model were very different to those in the original model, the code does not converge towards the correct solution for a significant number of pixels. This is the reason for repeating the inversion three times. With the spatial averaged values as input model in the second and third inversion, we help the code to approach to the correct solution for all pixels. The procedure helps to reach a better convergence although at the cost of computing time. It is important to remark that these tests have been carried out without taking into account the influence of noise sources coming from either the measurements process (i.e., telescope diffraction, limited sampling) or photon noise. This is beyond the scope of this paper and will be treated elsewhere. 

The results of the inversion will be compared with the original model atmosphere but in an optical depth scale. This model has been obtained during the spectral synthesis \citep{2018MNRAS.481.5675Q}. For completeness, the results obtained with mME code will also be compared with those obtained from the weak field approximation (WFA - Sect.~\ref{WFAsec}) for the magnetic field strength and inclination, and from the CoG technique (Sect.~\ref{CoGsec}) for the LOS velocity.

\subsection{Magnetic field strength}\label{mfsec}
  \begin{figure}[h]
  \centering
   \includegraphics[width=9cm]{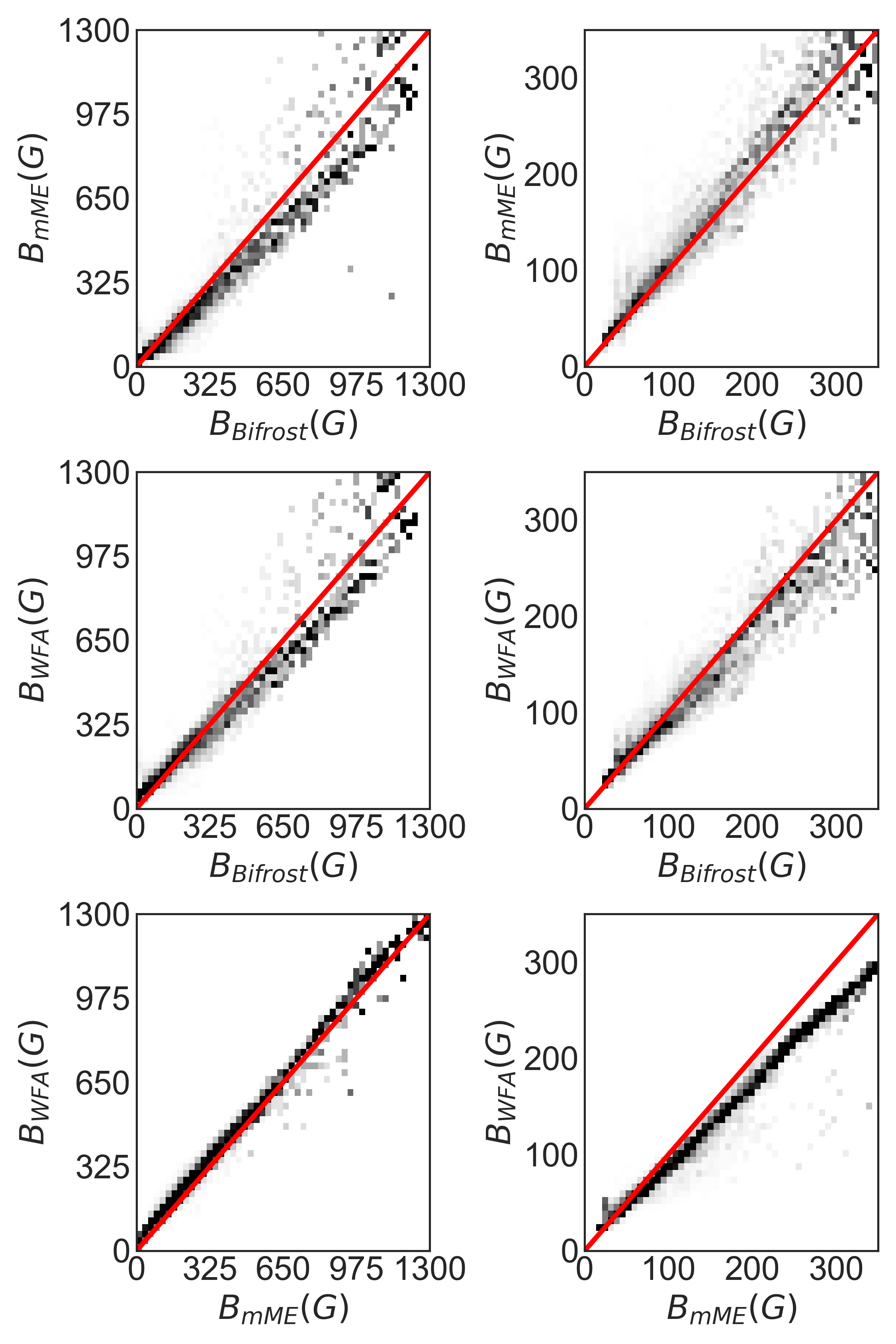}
      \caption{2D histograms of the magnetic field strength inferred from the Mg~{\small I}~b2 line (left column) and from the Ca~{\small II}~854.2 line (right column). Top panels show the results of mME versus Bifrost; middle panels show the results of the WFA versus Bifrost; and bottom panels correspond to the mME versus WFA.}
         \label{Hist}
   \end{figure}

Figure~\ref{BB} shows the maps of the magnetic field strength from the simulations at $\log\tau_{500}=-3.0$ and $\log\tau_{500}=-5.0$, optical depths that give the minimum rms when compared to the mME approximation applied to the Mg~{\small I}~b2 line and to the Ca~{\small II}~854.2 line, respectively. 
In terms of geometrical heights, these optical depths correspond to a wide range of kilometers whose averages correspond to $\sim 460$ km and $\sim 880$ km, respectively. 
 As expected, the Ca~{\small II}~854.2 provides information from higher layers than the Mg~{\small I}~b2 line. The figure also shows the maps of the field strength obtained from the mME inversion and from the WFA in the two lines. In the mME, magnetic field strength is one of the inverted physical parameters. In WFA we have to use transversal and longitudinal components - Eqs. (\ref{parB}) and (\ref{horB}) - to compute it with $B = \sqrt{B_\parallel^2+B_\perp^2}$. These components have been obtained by applying the approximation to the whole spectral range.
 
 The resemblance among the mME and WFA with the simulations is quite significant but looking into the details several mismatches can be appreciated. For instance, the inferred magnetic field maps show slightly stronger fields within magnetic field concentrations for both methods in the Mg~{\small I}~b2 case, while in the Ca~{\small II}~854.2 it only happens for mME. Likewise, the magnetic field inferred by the mME method and corresponding to the
 Ca line shows stronger fields than the WFA. Indeed, the rms are slightly smaller for the WFA than for the mME in this particular case. 

Figure~\ref{Hist} displays 2D histograms comparing  the magnetic field strengths resulting from the mME, WFA, and the Bifrost simulations. The dispersion is rather large for both, the mME and the WFA  when compared to Bifrost, which just highlights the fact that the inferences cannot be ascribed to single heights but the locations where the lines are sensitive to the magnetic field encompass a wide range of geometrical heights. The histograms in the bottom panels show a good correlation between mME and WFA for the Mg~{\small I}~b2 line; while for the Ca{\small II}~854.2 line the mME provides stronger fields than the WFA. This can also be seen, in Fig.~\ref{BB} where the WFA map shows weaker fields than the mME map, mainly in the regions with the strongest field concentrations. 

It is important to mention that the radiative transfer problem of the mME is a much stronger approximation than that of the WFA. This latter is limited to regions where the magnetic field is weak enough, or in detail, whenever the Zeeman splitting is much smaller than the Doppler broadening of the line. The good resemblance between the inferred magnetic field maps shown here indeed suggests that the WFA is adequate for having a glance at the longitudinal magnetic fields for chromospheric lines in weak field areas. However, it has to be kept in mind that the WFA will fail in those regions where the approximation does not hold (i.e., whenever Stokes $V$ is no longer proportional to the longitudinal component of the field). This can happen because of the underlying physics, e.g., strong field areas or instrumental effects such as the broadening of the line which induce linear deviations and hence a wrong inference of the longitudinal field. In this sense, we expect the mME approximation to provide more accurate results than the WFA, since it does not hold these limitations. 

\subsection{Magnetic field inclination}

    \begin{figure*}
   \centering
\includegraphics[width=19cm]{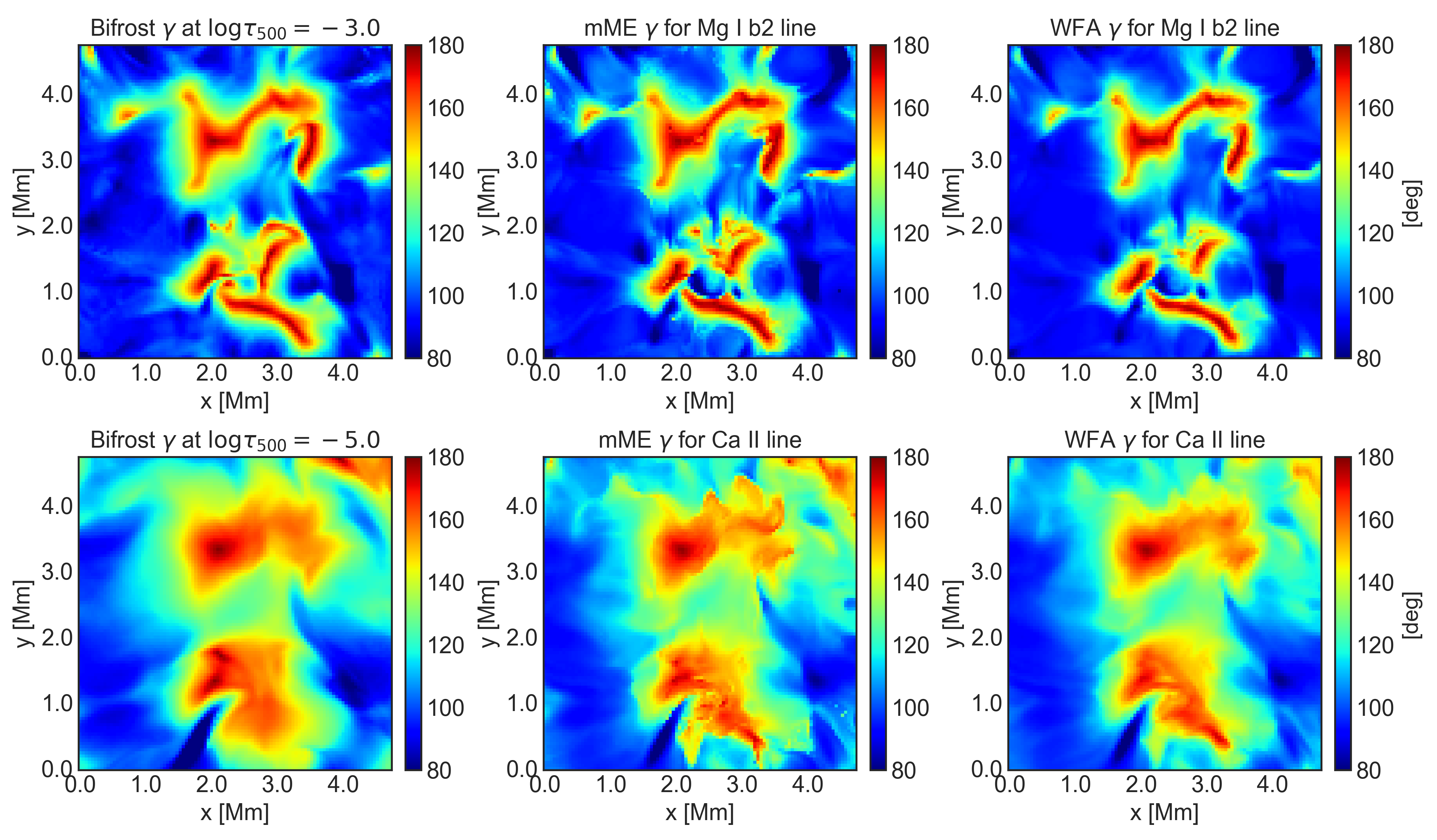}
      \caption{Magnetic field inclination ($\gamma$) maps of the region chosen for this study (in degrees). First row represents inclination corresponding to the Mg~{\small I}~b2 line. Left: Bifrost $\gamma$ at Mg~{\small I}~b2 formation height; Central: inverted $\gamma$ using mME; Right: inverted $\gamma$ using WFA. Second row contains inclination maps corresponding to Ca{\small II}~854.2 line. Left: Bifrost  $\gamma$ at Ca{\small II}~854.2 formation height; Central: inverted  $\gamma$ using mME; Right: inverted  $\gamma$ using WFA. Formation heights have been estimated choosing those whose residuals were minimum with the mME inverted B.}
         \label{gamma}
   \end{figure*}
   
 \begin{figure}[h]
  \centering
   \includegraphics[width=4.45cm]{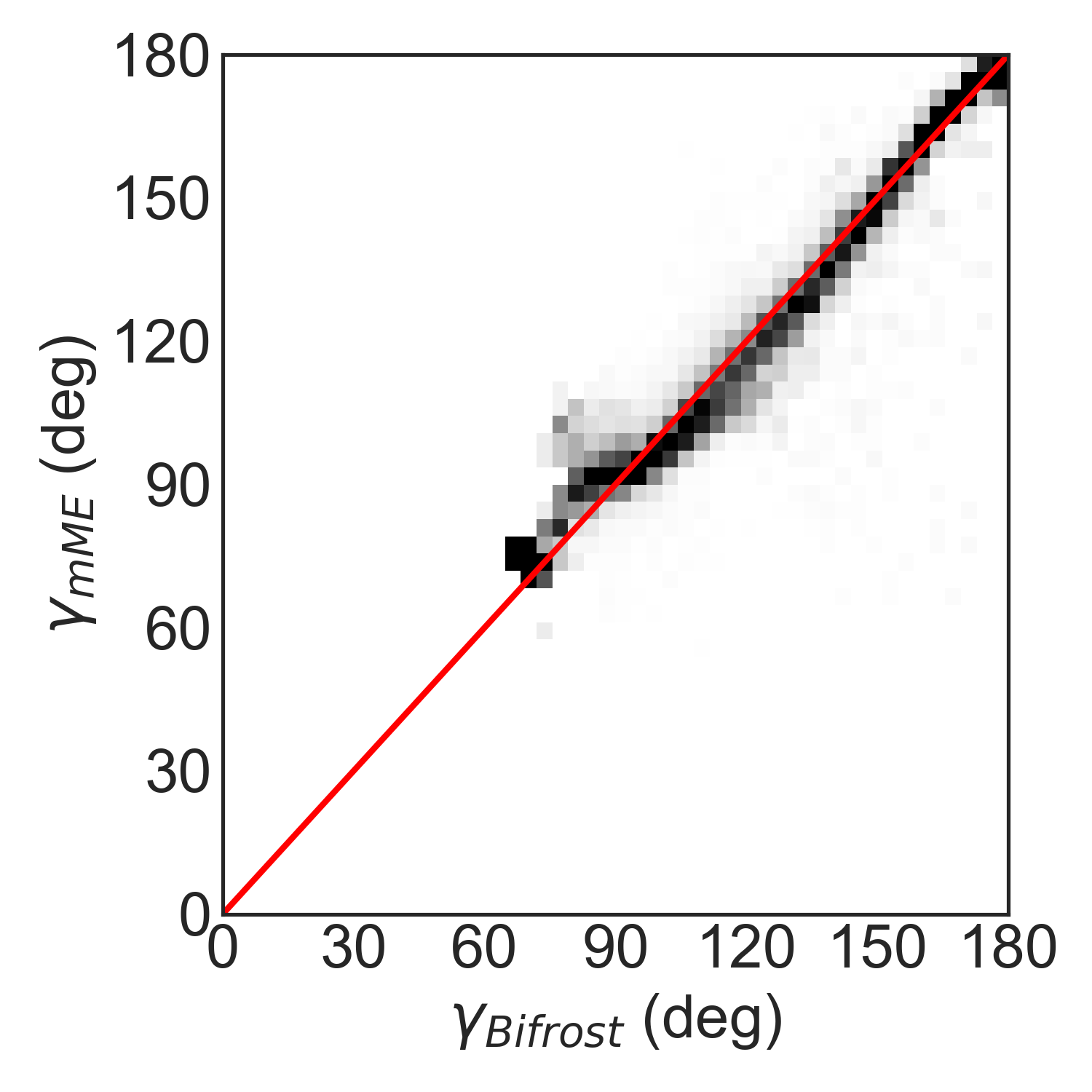}
   \includegraphics[width=4.45cm]{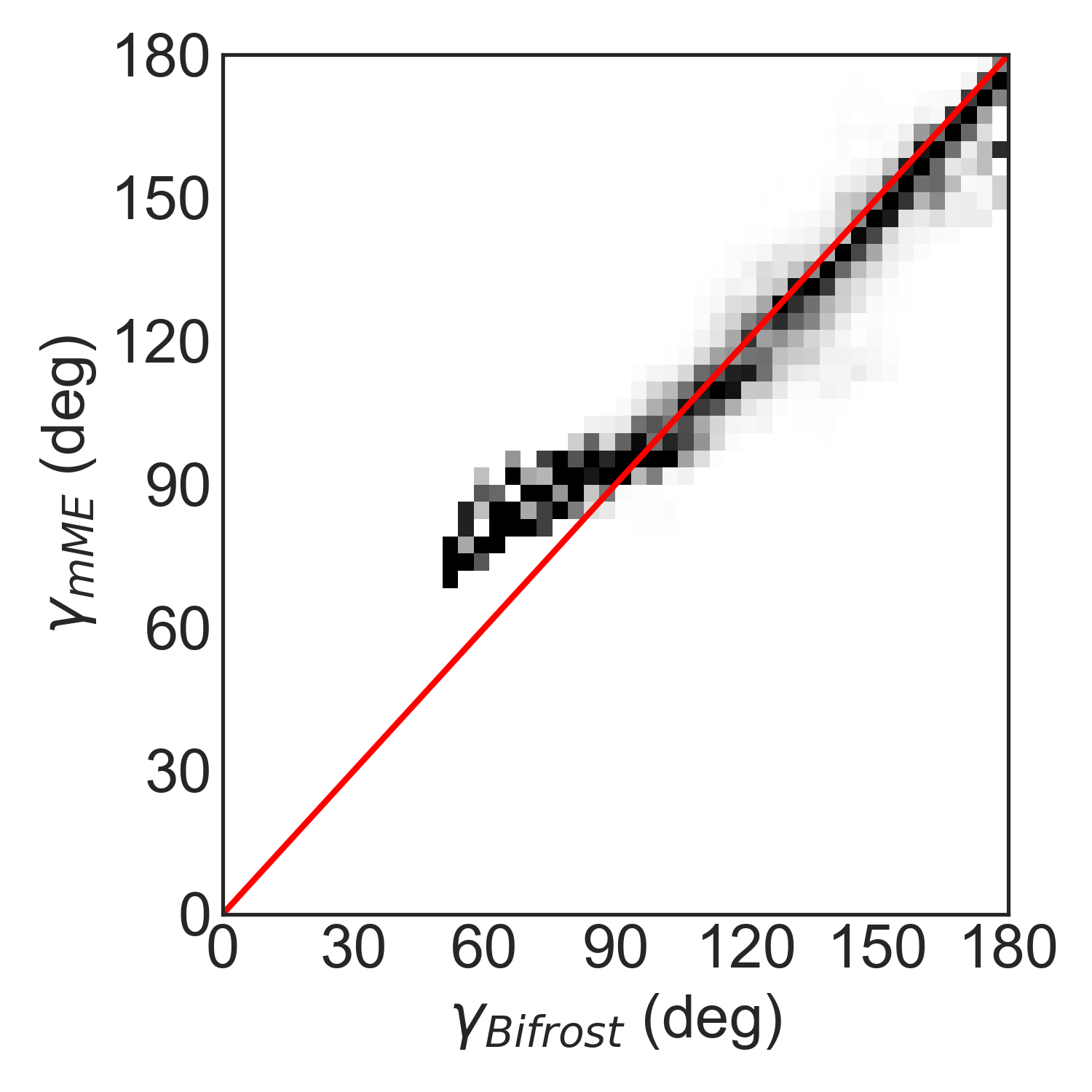}
      \caption{Normalized 2D inclination (in degrees) histogram for Mg~{\small I}~b2 line (left panel) and for Ca~{\small II}~584.2 line (right panel). We have represented mME inverted $\gamma$ vs. Bifrost $\gamma$ at line formation height. 
              }
         \label{incHist}
   \end{figure}

The results for the magnetic field inclination are shown in Fig.~\ref{gamma}. As for the magnetic field strength case, the figure shows the inclinations from the simulations taken at the same optical depths as in the field strength case, and those inferred with the mME approximation and the WFA, for both lines of interest. As we mentioned above, the magnetic field inclination ($\gamma$) is one of the physical quantities that we invert with the mME model atmosphere. However, for the WFA case, as we did with the magnetic field strength, the inclination has to be obtained from $B_\parallel$ and $B_\perp$, using Eqs.~(\ref{parB}) and (\ref{horB}) as:
   \begin{equation}
       \gamma = \arctan\left(\frac{B_\perp}{B_\parallel}\right) \label{WFAinc}
   \end{equation}

The magnetic structures appear more compacted in the maps of the Mg~{\small I}~b2 line  than in those for the Ca~{\small II} line. This is due to the opening of the magnetic field and the increase of the inclination with height.  The Mg~{\small I}~b2 show fields that are more vertical simply because this spectral line is sampling deeper layers.

Figure~\ref{incHist} shows histograms of mME inclination versus Bifrost's values for both lines. In both cases we can see a clear correlation between mME and the original atmospheres. There is a discrepancy for inclinations around 90 degrees (i.e., horizontal fields) in both lines. This is due to the fact that the majority of these horizontal fields are associated with weak fields in regions where the vertical stratifications change rather dramatically, hence, giving rise to strong asymmetries in Stokes $Q$ and $U$. This is the reason why the mME approximation fails in reproducing those profiles.

\subsection{Line-of-sight Velocities}
   \begin{figure*}[h]
   \centering
   \includegraphics[width=19cm]{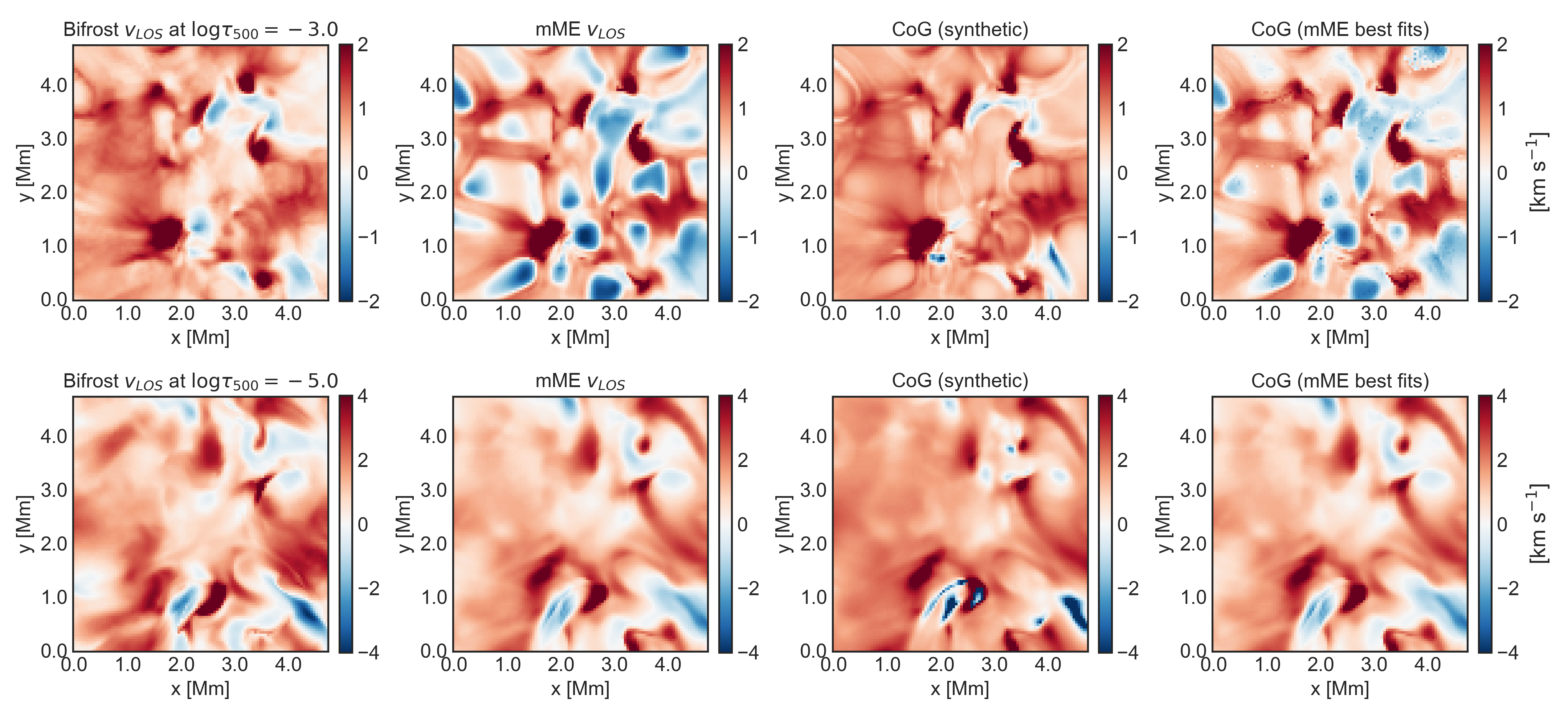}
      \caption{LOS velocity maps for the Mg~{\small I}~b2 line (top) and for the Ca~{\small II}~854.2 line (bottom). From left to right: Bifrost, mME inversions, CoG technique applied to the synthetic profiles, and CoG technique applied to the best fits obtained from the mME inversions.}
         \label{LOSv}
   \end{figure*}

Figure \ref{LOSv} shows LOS velocity maps for Mg~{\small I}~b2 and Ca~{\small II}~854.2 from the simulations, from the mME inversion, and from the CoG technique applied to the synthetic profiles and also applied to the best fit profiles resulting from the mME inversion.
The velocity maps for the Mg~{\small I}~b2 line from mME show a clear granulation pattern, typical of the photosphere, which is not seen at $\log\tau_{500}=-3.0$. In this case, the CoG applied to the synthetic profiles provides a more similar looking velocity map to Bifrost, while the CoG applied to the mME best fits also shows the granulation pattern as the mME inversions. The large differences might be due to large asymmetries in the Stokes $I$ profiles that are not taken into account by mME.

For Ca~{\small II} the velocity maps provide a more chromospheric appearance.
The larger structures seem to be well recovered but at small scales, there are clear differences between mME and the original maps. 
The results from the CoG also seem to be consistent with the original velocity map. The main difference is the presence of some small patches with strong blue-shifts that are not seen in the simulations.  

Finally, it can also be seen that the CoG applied to the best fits mME profiles retrieve, quite well, the inferred LoS velocity with the mME inversions, in both the Mg~{\small I}~b2 and the Ca~{\small II} lines, while the CoG applied to the synthetic profiles always display some differences with respect to mME. This result, although obvious, highlights the influence of spectral line asymmetries in the determination of the CoG wavelength within the line, asymmetries that are not taken into account by mME. 

\section{Conclusions} \label{conclusions}

In this paper we have presented a modification to the classical Milne-Eddington approximation which allows the interpretation of solar chromospheric lines, such as the Mg~{\small I}~b2 and the Ca~{\small II}~854.2~nm lines. The mME approximation was already tested by  \cite{1988ApJ...330..493L} using realistic chromospheric models although, in our opinion, not to great extent. Here, we have recycled the mME approximation since it allows for a \emph{faster} and simpler interpretation of chromospheric spectropolarimetric observations against those provided by full non-LTE inversions. The conclusions listed here are valid for both the Mg~{\small I}~b2 and the Ca~{\small II}~854.2~nm lines.
We have deepened in the usefulness and reliability of the mME approximation by first analyzing the so-called response functions under the mME approach. The analysis of the analytical mME response functions suggests that  trade-offs between the response functions to the vector magnetic field, the line-of-sight velocity, and the thermodynamic parameters are negligible. The latter is a necessary condition for an effective interpretation of chromospheric Stokes profiles under the mME approach since it allows inversion codes to disentangle the magnetic and velocity field information from the thermodynamic one. We have also shown that, at least, two exponential terms in the source function are needed for successfully reproducing chromospheric profile shapes unlike \cite{1988ApJ...330..493L} who concluded that only one exponential was sufficient. 
Next, we have tested its validity by applying it to synthetic profiles generated from a FALC model atmosphere including  different magnetic field configurations. From this analysis we conclude that:
\begin{enumerate}
    \item The mME approximation is able to reproduce chromospheric Stokes profiles emerging from standard atmospheric models and to determine the vector magnetic field rather accurately. The inferred mME model shows slight deviations in the determination of the field strength, for inclination values at around $90^\circ$ degrees.
    \item Although the mME source function shows similar behavior with optical depth as the ones computed from the original atmospheric model (i.e., it generally decreases monotonically with optical depth), it presents clear deviations. This fact  was not put forward in the work by \cite{1988ApJ...330..493L}. Particularly, the mME source function presents clear humps that cannot be found in the original one. The humps in the mME source function allow to fit the profile shape but have little physical meaning.
\end{enumerate}

After testing the mME approximation against the standard FALC model, we have confronted it with Stokes profiles generated from realistic MHD simulations  \citep{2016A&A...585A...4C, 2018MNRAS.481.5675Q}. For the sake of comparison, we have also used the WFA and the CoG techniques to compare the mME inversion results with simpler techniques. The results show that even though both, the mME and WFA, give similar results for the magnetic field strength and inclination, the mME interprets the data in a more comprehensive manner. For instance, the mME provides direct information about magnetic field strength, inclination or velocity fields unlike the WFA only provides independent measurements for the longitudinal and transverse components of the field vector. Another advantage of the mME approach is that it is not restricted to weak magnetic fields, unlike the WFA. 

\begin{acknowledgements}
This work has been funded by the Spanish Science Ministry of Science and Innovation through project RTI2018-096886-B-C51, including a percentage from FEDER funds, and through the Centro de Excelencia Severo Ochoa grant SEV-2017-0709 awarded to the Instituto de Astrof\'isica de Andaluc\'ia in the period 2018–2022. A.D.M aknowledges financial support through the Ph.D. grant BES-2017-082605 of the Ministry of Economy, Industry and Competitiveness. C.Q.N. was supported by the EST Project Office, funded by the Canary Islands Government (file SD 17/01) under a direct grant awarded to the IAC on ground of public interest, and this activity has also received funding from the European Union's Horizon 2020 research and innovation programme under grant agreement No 739500. D.O.S. acknowledges financial support through the Ram\'on y Cajal fellowship. 
\end{acknowledgements}

\end{document}